\RequirePackage[running]{lineno}

\documentclass[aps,prl,nofootinbib,superscriptaddress,letterpaper,amsmath,amssymb]{revtex4}
\pdfoutput=1

\usepackage{graphicx}  
\usepackage{dcolumn}   
\usepackage{bbm}        
\usepackage{amsmath}
\usepackage{amsfonts}
\usepackage{amssymb}   
\usepackage{epstopdf}
\usepackage{slashed}
\usepackage{hyperref}
\usepackage{multirow}
\usepackage{color}
\usepackage{float}
\usepackage{subcaption}
\usepackage{verbatim}

\usepackage[compat=1.1.0]{tikz-feynman}

\def\gsim{\lower0.5ex\hbox{$\:\buildrel >\over\sim\:$}}
\def\lsim{\lower0.5ex\hbox{$\:\buildrel <\over\sim\:$}}

\newcommand{\rev}[1]{{\color{black} #1}}

\newcommand{\be}{\begin{equation}}
\newcommand{\ee}{\end{equation}}
\newcommand{\bea}{\begin{eqnarray}}
\newcommand{\eea}{\end{eqnarray}}

\newcommand{\nbox}{{\,\lower0.9pt\vbox{\hrule \hbox{\vrule height 0.2 cm
\hskip 0.2 cm \vrule height 0.2 cm}\hrule}\,}}
 
 \newcommand{\lagr}{\mathcal{L}}

\def\sub#1{_{\lower.25ex\hbox{$\scriptstyle#1$}}}

\newskip\zatskip \zatskip=0pt plus0pt minus0pt
\def\matth{\mathsurround=0pt}
\def\lsim{\mathrel{\mathpalette\atversim<}}
\def\gsim{\mathrel{\mathpalette\atversim>}}
\def\sigv{\ifmmode \langle\sigma v\rangle\else $\langle\sigma v\rangle$\fi}
\newskip\zatskip \zatskip=0pt plus0pt minus0pt
\def\matth{\mathsurround=0pt}
\def\lsim{\mathrel{\mathpalette\atversim<}}
\def\gsim{\mathrel{\mathpalette\atversim>}}
\def\atversim#1#2{\lower0.7ex\vbox{\baselineskip\zatskip\lineskip\zatskip
  \lineskiplimit
  0pt\ialign{$\matth#1\hfil##\hfil$\crcr#2\crcr\sim\crcr}}}

\begin{document}

\thispagestyle{empty}
\vspace*{-3.5cm}

\vspace{0.5in}

\title{New Physics in Triboson Event Topologies}
\author{Linda M. Carpenter}
\author{Matthew J. Smylie}
\affiliation{Department of Physics, The Ohio State University}
\author{Jesus Manuel Caridad Ramirez}
\author{Cameron McDowell}
\author{Daniel Whiteson}
\affiliation{Department of Physics \& Astronomy, University of California, Irvine}
\begin{abstract}
We present a study of the sensitivity to models of new physics of proton collisions resulting in three electroweak bosons. As a benchmark, we analyze models in which an exotic scalar field $\phi$ is produced in association with a gauge boson ($V=\gamma$ or $Z$). The scalar then decays to a pair of bosons, giving the process $pp\rightarrow \phi V\rightarrow V'V''V$.  We interpret our results in a set of effective field theories where the exotic scalar fields couple to the Standard Model through pairs of electroweak gauge bosons.  We estimate the sensitivity of the  LHC and HL-LHC datasets and find sensitivity to cross sections in the 10 fb -- 0.5 fb range, corresponding to scalar masses of 500 GeV to 2 TeV and effective operator coefficients up to 35 TeV.

\end{abstract}
\maketitle

\section{Introduction}

Hadronic collisions at high energy are a powerful window into potential new particles and forces, whose existence may solve outstanding puzzles about the Standard Model or provide clues to new directions. The current run of the Large Hadron Collider (LHC),  however, has not yet revealed new physics beyond the Standard Model (BSM), despite extensive searching in many promising production and decay modes.  But significant opportunities remain in unexamined event topologies, including asymmetric two-body decay modes~\cite{Craig:2016rqv,Kim:2019rhy} as well as two-step decays~\cite{Agashe:2017wss} which lead to three or more objects in the final state.

As the LHC dataset grows large,  opportunities are created for the study of rare final states, such as those with three weak vector bosons ($V$), allowing for new tests of the Standard Model and searches for physics beyond it.  The ATLAS and CMS Collaborations have recently reported observation of $VVV$ production with $V = W,Z$ in events with two to six leptons \cite{ATLAS:2019dny,CMS:2020hjs}. This measurement may help to extract the value of sensitive Standard Model parameters \cite{Falkowski:2020znk}. The triboson channel may also be a powerful window to Beyond the Standard Model scenarios; if two of the bosons in a triboson event are due to the decay of a new heavy particle, $\phi\rightarrow VV$, a resonance peak in the diboson mass spectrum may be a clear discovery signature. While there are several dedicated searches for diboson resonances~\cite{ATLAS:2017jag,ATLAS:2017zuf,CMS:2017fgc,CMS:2018ygj,ATLAS:2019nat}, they typically consider only the leading two reconstructed vector bosons, and do not develop dedicated algorithms to search for $VV$ resonances among the full $VVV$ triplet.

In this paper, we study the sensitivity of proton collisions which result in three electroweak bosons to a benchmark model of new physics containing a new heavy scalar $\phi$, produced in association with a boson $V$, and which decays to a pair of bosons, $\phi\rightarrow V'V''$, giving the process $pp\rightarrow \phi V\rightarrow V+V'V''$.   The associated production and subsequent decay are enabled by the new scalar's diboson coupling;  in fact the  diboson  coupling serves as the portal between an exotic sector and the Standard Model in several interesting BSM scenarios, such as Higgs imposter fields \cite{Low:2011gn,Carpenter:2012zr}, and Dark Matter models \cite{Carpenter:2016thc,Carpenter:2015xaa,Carpenter:2012rg,Lopez:2014qja,Nelson:2013pqa}. In order to discuss scalar models as generally as possible, we construct a set of simple Effective Field Theories (EFTs), in which new spin-zero states in simple representations of Standard Model gauge groups couple to pairs of Standard Model gauge bosons.

We focus on fully-reconstructable decays, which allow for identification of a sharp resonance peak, and expand the triboson searches to include photons, exploring the $V=\gamma$ and $V=Z$ scenarios.  We leave $V=W$ and final states with neutrinos to future studies, some of which are already explored in Ref.~\cite{Agashe:2017wss}.  The many combinations of $Z$ and $\gamma$ production and their large number of decay modes make for a rich phenonemology; we present studies of 14 distinct final states, which have the greatest sensitivity to our benchmark model.  We find that the LHC dataset is sensitive to these processes at cross sections of $0.5-10$ fb for new states with masses $0.5-2$ TeV, corresponding to bounds on effective mass scales of EFT operators ranging from $0.5-35$ TeV.

In Section 2, we discuss effective field theory models, and in Section 3 we describe experimental signatures for the triple electroweak gauge boson signature. Section 4 details the calculation of experimental sensitivities to each signature, and Section 5 presents expected limits in terms of the EFT scales. 

\section{Models}
\label{sec:model}

Diboson couplings present a potential portal into an exotic scalar sector, which appears in many  BSM theories. For example, the sgaugino sector of R-symmetric SUSY models \cite{Fox:2002bu} contains a large family of new scalar and pseudo-scalar fields charged under SM  gauge groups,  which may have their main decay modes through loop-level diboson couplings \cite{Choi:2010gc,Carpenter:2021tnq,Carpenter:2020evo}. Various exotic scalars have also been invoked to explain possible diboson resonance signatures~\cite{McDermott:2015sck,Low:2015qep,Angelescu:2015uiz,Carpenter:2015ucu,Aguilar-Saavedra:2015iew}.

Due to the broad set of UV theories which in which new scalars may arise, we construct EFTs which allow access to the generalized weak-scale phenomenology without being sensitive to the UV details.  We construct a detailed catalogue of effective operators up to dimension 7 which couple exotic states to pairs of Standard Model gauge bosons. In total generality, such a list is quite daunting, and even with theoretical simplifications the collider phenomenology of such a roster of operators is very complex. For the sake of simplicity, we here only consider exotic spin-zero fields in the singlet, fundamental, and adjoint representations of the Standard Model gauge groups. 

Below we list all such operators which are gauge and Lorentz invariant. Each effective operator will be suppressed by a new physics scale $\Lambda$. One general feature of the operators is that additional Higgs fields may be included to soak up extra $SU(2)$ indices at the cost of raising the dimension of the operators. Once Higgs vevs ($v$) are inserted, the effective dimension of these operators will decrease,  paying a price of powers of a scale factor $v/\Lambda$.
We note in particular that the electroweak associated production channel allows us to study the production of a single exotic particle that does not couple (or couples very weakly) to gluons and therefore does not allow a $pp\rightarrow \phi$ production mode. We will specialize to this case in our interpretations in Section 5.

We begin by considering a total SM singlet scalar $X$. We give the SM charges in the table below, followed by the lowest dimension set of effective operators which couple this field to pairs of gauge bosons.

\begin{table}[H] 
\centering 
\begin{tabular}{|c|c|c|c|} 
\hline Field & $U(1)_Y$ & $SU(2)$ & $SU(3)$ \\ \hline
$X$ & 0 &  1 & 1 \\
\hline
\end{tabular} 
\end{table}

\begin{equation}
\mathcal{L}_1= \frac{1}{\Lambda_{XBB}} X B^{\mu \nu} B_{ \mu \nu}+ \frac{1}{\Lambda_{XWW}} X W^{\mu \nu} W_{ \mu \nu} + \frac{1}{\Lambda_{XGG}} X G^{\mu \nu} G_{ \mu \nu}+\frac{1}{\Lambda_{XBW}^3}  X B^{\mu \nu} [H^{\dagger}W_{ \mu \nu} H] 
\end{equation}

The first three operators are dimension 5 with the scalar $X$ coupling to pairs of SM field strength tensors. The last operator above couples the singlet $X$ to the $SU(2)$ and $U(1)$ field strength tensors. In this operator, extra $SU(2)$ indices are contracted with Higgs fundamental and anti-fundamentals; here and below, square brackets are used to denote full contraction of $SU(2)$ indices. The operator is dimension 7, but the two inserted Higgs vevs effectively bring it to dimension 5.  The first two operators  couple  the singlet $X$ into four distinct pairs of electroweak bosons $ZZ, WW, Z\gamma$ and $\gamma \gamma$. The third operator couples $X$ to pairs of gluons, and the last couples the neutral $X$ to $ZZ, Z\gamma$ and $\gamma\gamma$ pairs.

Next, we  consider a scalar state with  Higgs-like SM gauge indices, an $SU(2)$ doublet with hypercharge 1.

\begin{table}[H] 
\centering 
\begin{tabular}{|c| c | c |c|} 
\hline Field 
& $U(1)_Y$ & $SU(2)$ & $SU(3)$ \\
\hline
$Y$ & 1 &  2 & 1 \\ 
\hline
\end{tabular} 
\end{table}

\begin{equation} 
\mathcal{L}_2= \frac{1}{\Lambda_{YBB}^2} [H^{\dagger}  Y] B^{\mu \nu} B_{ \mu \nu}+ \frac{1}{\Lambda_{YWW}^2} [H^{\dagger}  Y] W^{\mu \nu} W_{ \mu \nu} + \frac{1}{\Lambda_{YGG}^2} [H^{\dagger}  Y] G^{\mu \nu} G_{ \mu \nu}+\frac{1}{\Lambda_{YBW}^2}   B^{\mu \nu} [H^{\dagger}W_{ \mu \nu} Y] 
\end{equation}

In the first three operators above, we build the bi-linear $H^{\dagger} Y$. This bi-linear is again a total SM singlet and may be thus coupled to pairs of the SM field strength tensor at dimension 6. Once the Higgs vevs are inserted these operators become effective dimension 5. The final operator is again of dimension 6, and in this term the $SU(2)$ indices are contracted between one Higgs doublet, the $SU(2)$ field strength tensor and the new field $Y$. Again once the Higgs vev is inserted this becomes an effective dimension 5 operator. These operators are closely related to the ones above and have the same pattern of couplings to gauge boson mass eignenstates.

We now consider fields in the adjoint representation of SM gauge groups. We define an $SU(2)$ triplet field $T$ and an $SU(3)$  octet field $O$. The lowest dimension operators coupling these adjoint fields to SM gauge bosons are dimension 5, and we write them here:

\begin{table}[H]  
\centering 
\begin{tabular}{|c|c|c|c|} 
\hline Field 
& $U(1)_Y$ & $SU(2)$ & $SU(3)$ \\
\hline
$T$ & 0 & 3 & 1 \\
\hline
$O$ & 0 & 1 & 8 \\ 
\hline 
\end{tabular} 
\end{table}

\begin{equation} 
\mathcal{L}_3= \frac{d^{abc}}{\Lambda_{2}} O_a G_b^{\mu \nu} G_{c, \mu \nu} + \frac{1}{\Lambda_{1}} O_{a} G^{a,\mu \nu} B_{ \mu \nu} + \frac{1}{\Lambda_{TWB}} T_i W_i^{\mu \nu} B_{ \mu \nu} 
\end{equation}

The first two operators above involve the octet $O$.  The first operator couples the octet to pairs of gluons; color indices are here contracted symmetrically between the octet and the two $SU(3)$  field strength tensors. The second operator couples an octet to the $SU(3)$  and $U(1)$ field strength tensors, with $SU(3)$  indices contracted between the octet and the field strength tensor. As studied in reference \cite{Carpenter:2015gua}, these two operators couple the octet to gluon-gluon, gluon-photon and gluon-$Z$ pairs of bosons. The last operator involves the $SU(2)$ triplet scalar which couples to the $SU(2)$ and $U(1)$ field strength tensors. Here $SU(2)$ indices are contracted between the triplet scalar and the $SU(2)$ field strength tensor.  This operator couples the neutral component of the triplet to photon-photon, photon-$Z$ and $ZZ$ pairs.

We may also consider higher dimensional operators which couple these adjoints to pairs of SM gauge bosons. There is a set of dimension 7 operators in which $SU(2)$ indices are soaked up with two Higgs contractions. We write them below:

\begin{equation} 
\mathcal{L}_4= \frac{1}{\Lambda_{TBB}^3} [H^{\dagger} T H] B^{\mu \nu} B_{ \mu \nu}+ \frac{1}{\Lambda_{TWW}^3} [H^{\dagger} T H] W^{\mu \nu} W_{ \mu \nu} + \frac{1}{\Lambda_{TGG}^3} [H^{\dagger} T H] G^{\mu \nu} G_{ \mu \nu}+\frac{1}{\Lambda_{OGW}^3}  O^a G_a^{\mu \nu} [H^{\dagger}W_{ \mu \nu} H] 
\end{equation}

The first three operators above involve the triplet $T$. Here we construct an $SU(2)$ singlet by contracting the triplet with two Higgs fields. Thus the operators coupling $T$ to pairs of gauge bosons are dimension 7, but they become effective dimension 5 once Higgs vevs are inserted. The first two operators will couple the charged and neutral components of $T$ to pairs of electroweak gauge bosons while the third operator will couple the neutral component of $T$ to pairs of gluons. The last operator involves the $SU(3)$  octet $O$ and couples the $SU(3)$  and $SU(2)$ field strength tensors. Here $SU(3)$  indices are contracted between the octet and the $SU(3)$  field strength tensor, while $SU(2)$ indices are contracted between the $SU(2)$ field strength tensor and two Higgs insertions.  The octet can then decay to gluon-$Z$ and gluon-photon pairs.

Next we consider fields with both $SU(2)$ and $SU(3)$ indices, beginning with a color octet field that is also a fundamental under $SU(2)$ as in the Manohar-Wise model \cite{Manohar:2006ga}. We add two such fields $S_u$ and $S_d$, in analogy with two Higgs doublet models. In order to have integer charge, we give these fields hypercharges of $\pm 1$. These states may couple to pairs of gauge bosons with operators of dimension 6 as listed below.

\begin{table}[H] 
\centering 
\begin{tabular}{|c| c | c |c|} 
\hline Field 
& $U(1)_Y$ & $SU(2)$ & $SU(3)$ \\ 
\hline
$S_u$ & 1 &  2 & 8 \\ 
\hline 
$S_d$ & -1 & 2 & 8 \\
\hline
\end{tabular} 
\end{table}

\begin{equation} 
\begin{split} \mathcal{L}_5&= \frac{1}{\Lambda_{gg1}^2}d^{abc} [H^{\dagger} S_{ua}] G_{b}^{\mu \nu}G_{c}^{\mu \nu}+ \frac{1}{\Lambda_{gg2}^2} d^{abc}[H S_{da}] G_{b}^{\mu \nu}G_{c}^{\mu \nu} + \frac{1}{\Lambda_{gb1}^2} [H^{\dagger} S^a_u] G_{a}^{\mu \nu}B^{\mu \nu}+ \frac{1}{\Lambda_{gb2}^2} [HS_d^a] G_{a}^{\mu \nu}B^{\mu \nu} \\  &+ \frac{1}{\Lambda_{gw1}^2} [H^{\dagger}W^{\mu \nu}S_u^a] G_{a}^{\mu \nu}+ \frac{1}{\Lambda_{gw2}^2} [H W^{\mu \nu}S_d^a] G_{a}^{\mu \nu} \end{split} 
\end{equation}

In the first four operators, $SU(2)$ indices are contracted between the states $S_u$ or $S_d$ and the Higgs field $H$ in a bilinear term. In the first two, the remaining color index is then contracted symmetrically with two $SU(3)$  gauge field strength tensors, and in the second two, the bilinear term is contracted with a single $SU(3)$ field strength tensor and the $U(1)$ field strength tensor. In the final two operators, the $SU(2)$ structure is a bit different. Here an $SU(2)$ singlet is constructed by contracting a BSM doublet, the Higgs doublet, and the $SU(2)$ field strength tensor.  $SU(3)$ indices are then contracted between this trilinear and the $SU(3)$  field strength tensor. These operators couple the neutral part of $S_u$ and $S_d$ to gluon-gluon, gluon-$Z$ and gluon-photon pairs.

Finally we consider a field which is an adjoint both under $SU(2)$ and $SU(3)$. This field may be coupled to pairs of electroweak bosons through dimension 5 or 7 operators as shown below.

\begin{table}[H]  
\centering
\begin{tabular}{|c| c | c |c|} 
\hline Field 
& $U(1)_Y$ & $SU(2)$ & $SU(3)$ \\
\hline
$S$ & 0 &  3 & 8 \\ 
\hline 
\end{tabular} 
\end{table}

\begin{equation}
\mathcal{L}_6= \frac{1}{\Lambda_{sgw}} S_{a}^{i} G_{a}^{\mu \nu} W_{\mu \nu}^{i} + \frac{1}{\Lambda_{sgb}^3} [H^{\dagger} S_{a}^{}H] G_{a}^{\mu \nu} B_{\mu \nu}^{} + \frac{1}{\Lambda_{sgg}^3} d^{abc}  [H^{\dagger} S_{a}^{}H] G_{b}^{\mu \nu} G_{c\ \mu \nu}^{} \end{equation}

The first operator is a dimension 5 coupling between the bi-adjoint and the $SU(2)$ and $SU(3)$  field strength tensors. The electrically neutral scalar component will thus couple to gluon-photon and gluon-$Z$ pairs while the charged component will couple to $W$-gluon pairs. The last two operators are dimension 7 operators which become effective dimension five when Higgs vevs are inserted. In these operators $SU(2)$ indices are contracted in a trilinear term between the bi-adjoint and two Higgs fields. In the second term $SU(3)$  indices are contracted with the $SU(3)$  field strength tensor. In the third term $SU(3)$  indices of the trilinear are contracted symmetrically with two $SU(3)$  field strength tensors.  These last two operators couple the bi-adjoint to gluon-gluon, gluon-photon, and gluon-$Z$ pairs of gauge bosons.

These operators lead to various production modes of scalar states, gluon fusion, vector boson fusion, and vector boson associated production.
The correspondence between production modes and operators is given in
Table \ref{process-table}.
\begin{table}[h]
\centering
\begin{tabular}{|c|c|}
\hline 
Process & Operator\tabularnewline
\hline 
\hline 
gluon fusion & 1.3 , 2.3, 3.3, 4.3, 5.1, 5.2, 6.3\tabularnewline
\hline 
Vector Boson Fusion & 1.2, 1.2, 1.4, 2.1, 2.2, 2.4, 3.2, 3.3, 4.1, 4.2, 4.4, 5.3-5.6, 6.1,
6.2\tabularnewline
\hline 
associated production, electroweak & 1.1, 1.2, 1.4, 2.1, 2.2, 2.4, 3.2, 3.3, 4,1, 4.2, 4.4, 5.3-5.6, 6.1,
6.2 \tabularnewline
\hline 
associated production, gluon & 1.3, 2.3, 3.1, 3.2, 4.3, 4.4, 5.1-5.6, 6.1-6.3\tabularnewline
\hline 
\end{tabular}
\caption{The production modes of scalar states and their correspondence to the operators of the models described above. In this table, the operator $m.n$ should be understood as the $n$th term of $\lagr_m$.}
\label{process-table}
\end{table}

In each production mode, we consider that the new exotic states are produced and will decay to two vector bosons. There are a large variety of interesting final states which result; among the most striking, however, are those involving the production of three gauge bosons. Moreover many operators listed above contribute to final state collider topologies in which there are three electroweak gauge bosons. We will next discuss triple electroweak boson signatures as a striking search strategy for new physics.

\section{Triple Electroweak Gauge Boson Signature }

\begin{figure}[h!]
	\centering
	\scalebox{1.3}{
		\begin{tikzpicture}
			\begin{feynman}
				\vertex (a);
				\vertex [above left=of a](i1){\(q\)};
				\vertex [below left=of a](i2){\(q\)};
				\vertex [right=2cm of a,blob](b){};
				\vertex [above right=2cm of b,blob](c){};
				\vertex [below right=2cm of b](f1){\(V\)};
				\vertex [right=of c](g);
				\vertex [above=0.5cm of g](f2){\(V'\)};
				\vertex [below=0.5cm of g](f3){\(V''\)};
				
				\diagram*{
					(i1) --[fermion] (a),
					(i2) --[anti fermion] (a),
					(a) --[boson] (b),
					(b) --[scalar,edge label=\(\phi\)] (c),
					(b) --[boson] (f1),
					(c) --[boson] (f2),
					(c) --[boson] (f3),
				};
			\end{feynman}
		\end{tikzpicture}
	}
	\caption{Production of the scalar state $\phi$ in association with an electroweak boson $V$, with subsequent decay $\phi\rightarrow VV$ giving a three-boson final state.}
	\label{fig:decay-diagram}
\end{figure}
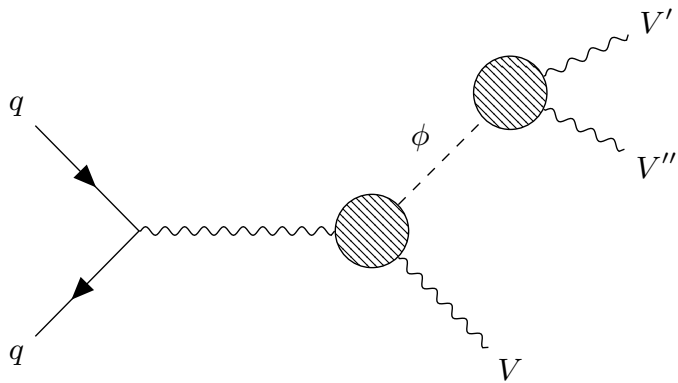

The models described above include interactions which can generate a final state with three electoweak bosons ($V$), via the  production of an exotic particle $\phi$ in
association with one electroweak gauge boson $ q q\rightarrow \phi V$, with subsequent decay $\phi\rightarrow VV$; see Fig.~\ref{fig:decay-diagram}.

To leverage the power of LHC experiments to reconstruct narrow resonances, we focus on signatures which include no missing momentum via the production of neutrinos. To reduce the number of combinations, we further restrict ourselves to $V=\gamma$ and $Z$; see Ref.~\cite{Agashe:2017wss} for discussion of some modes with missing momentum and involving $W$ bosons.

\rev{The couplings of the novel scalar states to weak bosons  also allows them to be produced via vector boson fusion (VBF). However, the production cross section for this mode turns out to be smaller than that of associated production, with larger expected backgrounds due to the two forward jets which accompany VBF. These factors make VBF less powerful for constraining our models. Therefore, we specifically focus only on associated production in our analyses below.}

We consider hadronic and charged-leptonic decay modes of the $Z$, excluding $\tau$-leptons. The multi-stage decay process results in one, two, or three reconstructable resonances.  To indicate which objects are due to a resonance at a previous stage, we use parentheses, as in $\phi V \rightarrow (V'V'')V$, to indicate that $V'$ and $V''$ are due to the $\phi$ decay, whereas $V$ is produced in association.   We consider all possible decays of the $Z$, and below present the 14 production and decay modes which provide the most powerful constraints on the model; see Table~\ref{tab:modes}.

\begin{table}[]
    \centering
    \begin{tabular}{l|l}
            \hline        \hline
        Production and decay & Final state  \\
        \hline
         $\phi \gamma \rightarrow (\gamma \gamma) \gamma$ & $3\gamma$ \\ \hline
$\phi \gamma \rightarrow (ZZ) \gamma \rightarrow ( J (ee) ) \gamma$  & $J, 2\ell,\gamma$\\
$\phi \gamma \rightarrow (ZZ) \gamma \rightarrow ( J (\mu\mu) ) \gamma$ & \\ \hline
$\phi \gamma \rightarrow (ZZ) \gamma \rightarrow ( (ee) (ee) ) \gamma$ & $4\ell, \gamma$\\
$\phi \gamma \rightarrow (ZZ) \gamma \rightarrow ( (ee) (\mu\mu) ) \gamma$& \\
$\phi \gamma \rightarrow (ZZ) \gamma \rightarrow ( (\mu\mu) (\mu\mu) ) \gamma$& \\
$\phi Z \rightarrow (Z \gamma) Z \rightarrow ( (ee) \gamma ) (ee)$ & \\
$\phi Z \rightarrow (Z \gamma) Z \rightarrow ( (ee) \gamma ) (\mu\mu)$& \\
$\phi Z \rightarrow (Z \gamma) Z \rightarrow ( (\mu\mu) \gamma ) (ee)$& \\
$\phi Z \rightarrow (Z \gamma) Z \rightarrow ( (\mu\mu) \gamma ) (\mu\mu)$& \\ \hline
$\phi \gamma \rightarrow (\gamma Z) \gamma \rightarrow ( \gamma (\mu\mu) ) \gamma$ & $2\ell,2\gamma$\\
$\phi \gamma \rightarrow (\gamma Z) \gamma \rightarrow ( \gamma (ee) ) \gamma$ & \\
$\phi Z \rightarrow (\gamma \gamma) (e e)$ & \\
$\phi Z \rightarrow (\gamma \gamma) (\mu \mu)$ &\\
            \hline        \hline
    \end{tabular}
    \caption{ Production and decay modes considered in this study, where parenthesis indicate reconstructable resonances. Also indicated are the elements of the experimental final state, where $J$ refers to a large-radius jet from a $Z\rightarrow qq$ decay, and $\ell$ is an electron or muon.}
    \label{tab:modes}
\end{table}

\section{Experimental Sensitivity}

We estimate the sensitivity of the LHC dataset to these hypothetical signals using samples of simulated collisions representing 100 fb$^{-1}$ of proton collisions.

\subsection{Simulated Samples}

Simulated signal and background samples were used to reconstruct hypothetical resonances, estimate efficiencies and expected yields.
Collisions and  decays are simulated with {\sc Madgraph5} v2.9.2 ~\cite{madgraph}, showered and hadronized with {\sc Pythia} v8.235~\cite{pythia}, and the detector response is simulated with {\sc Delphes} v3.4.1~\cite{delphes} using the standard ATLAS card and {\sc root} version 6.08\/00 \cite{ROOT}.  \rev{The theoretical coupling constants were set to default values of $10^{-4}\ \mathrm{GeV}^{-1}$, and the cross-sections for a particular production and decay mode are calculated by multiplying the leading-order cross-section by the branching ratios relevant to the decay chain. As is typically done, we assume that the reconstructed invariant mass distribution of the new states are dominated by experimental resolution and so independent of the values of the coupling.}

Selected photons and leptons are required to have transverse momentum $p_\textrm{T}\geq10$ GeV and absolute pseudo-rapidity $0\leq|\eta|\leq2.5$. Selected jets are clustered using the anti-$k_{\textrm{T}}$ algorithm~\cite{Cacciari:2008gp} with radius parameter $R = 0.8$ using \textsc{FastJet 3.1.2}~\cite{Cacciari:2011ma} and are required to have $p_\textrm{T}\geq20$ GeV and $0\leq|\eta|\leq5$.

\subsection{$\gamma\gamma\gamma$ final state}

In this section, we consider the $\phi \gamma \rightarrow (\gamma \gamma) \gamma$ mode, which produces a $3\gamma$ final state. The selection  requires at least three photons. The major background process is Standard Model (SM) $\gamma\gamma\gamma$ production without the $\phi$ resonance; secondary backgrounds, such as those in which a jet is misidentified as a lepton, are not considered here. From the selected photons, the $\phi$ candidate is reconstructed from the pair of photons with the largest $\Delta{}p_\textrm{T}$, which gives the narrowest reconstructed resonance in our studies. 

\begin{figure}[h!]
    \centering
    \scalebox{0.4}{\includegraphics{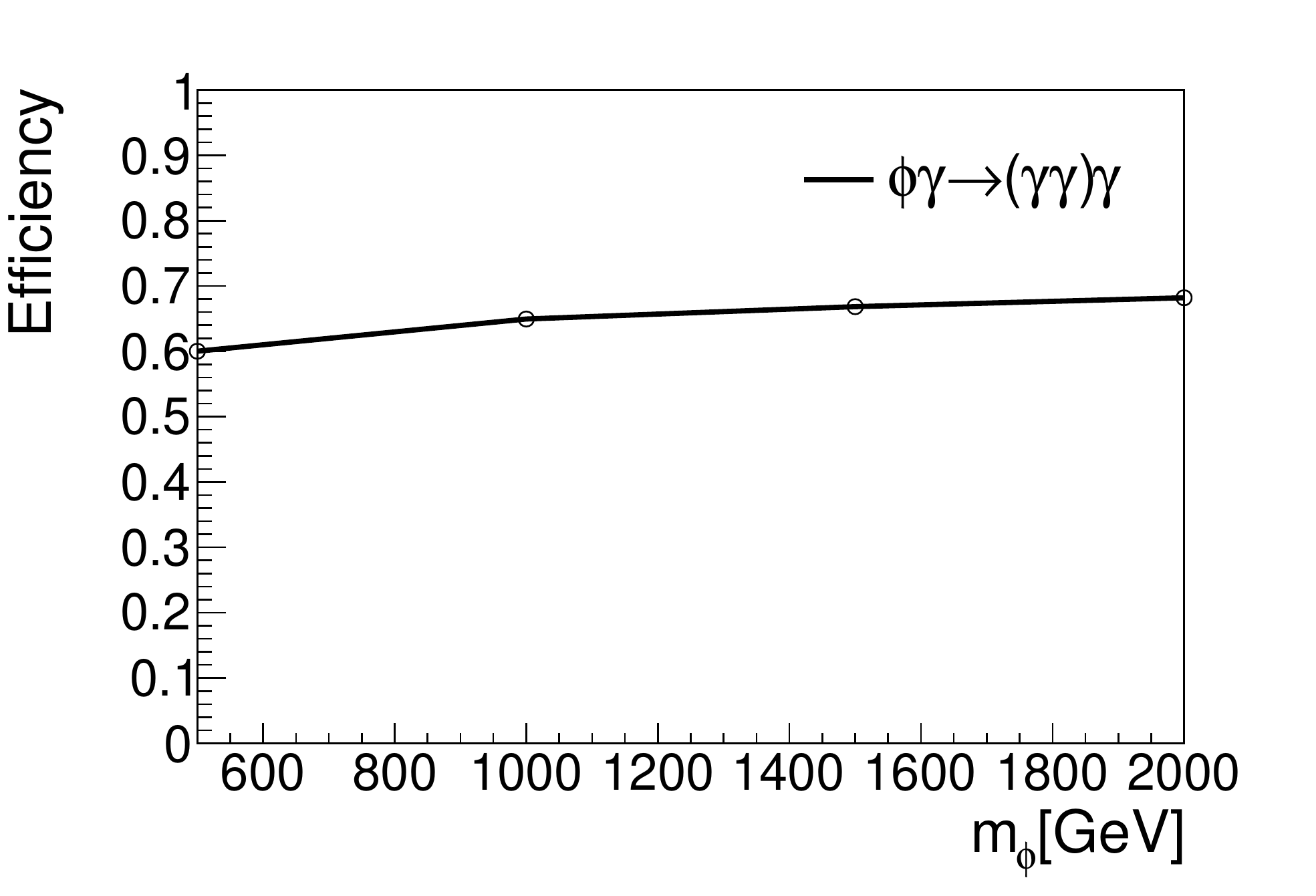}}
    \scalebox{0.4}{\includegraphics{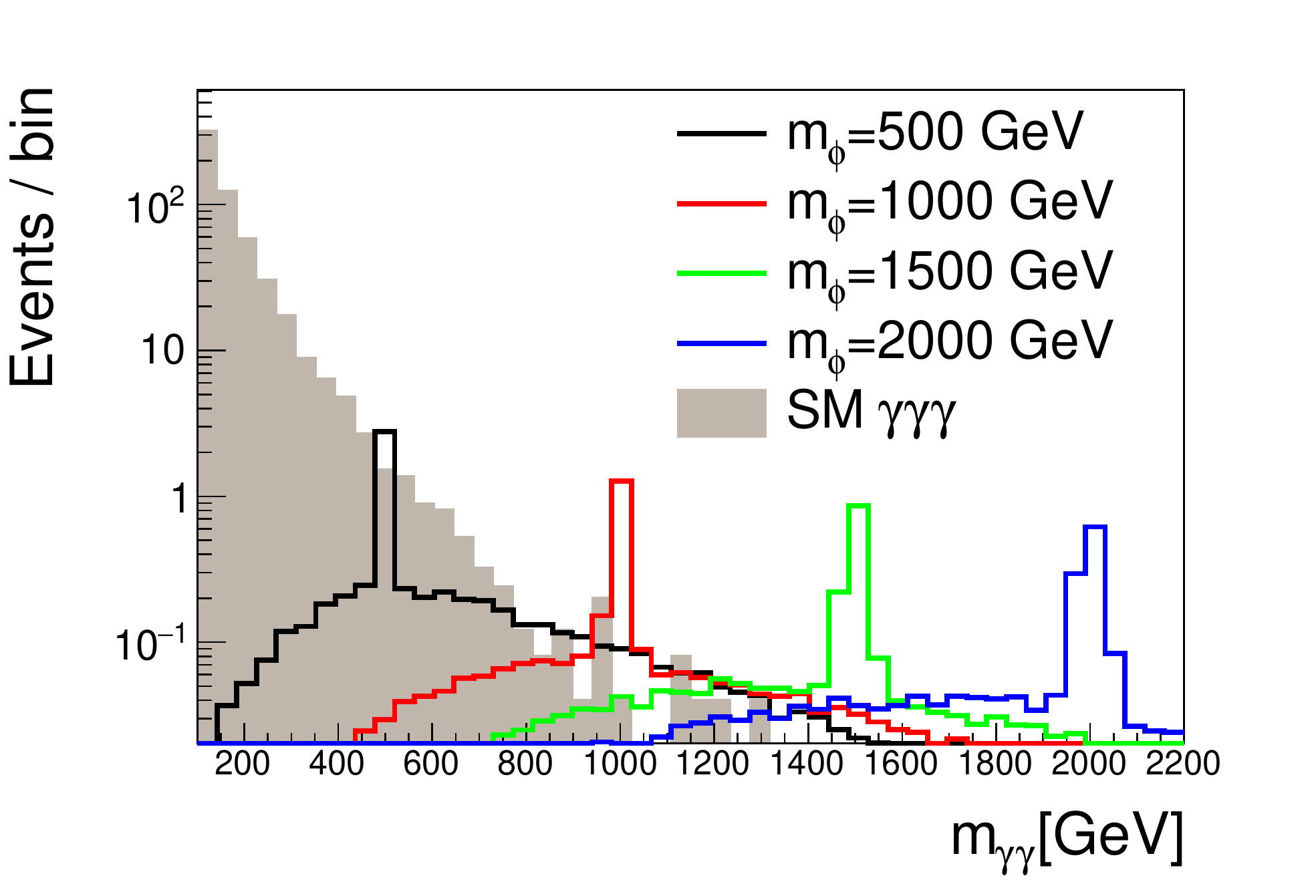}}
    \caption{(Left) Efficiency of three-photon selection as a function of the hypothetical $\phi$ mass for the $\gamma\gamma\gamma$ final state. (Right) Distributions of reconstructed $m_\phi$ in simulated signal and background samples, normalized to integrated luminosity of 100 fb$^{-1}$. \rev{The signal cross section is calculated at leading order assuming the default values of the coupling constants; see text for details.}}
    \label{fig:3gamma_plots}
\end{figure}

The efficiency of the selection versus $\phi$ mass, and the distributions of reconstructed $m_\phi$ for signal and background samples are shown in in Figure~\ref{fig:3gamma_plots}. The SM background falls smoothly and rapidly. The signal is very narrowly peaked due to the excellent photon energy resolution, with broader tails in cases where the correct pair of photons are not selected to form the $\phi$ candidate.

\subsection{$\gamma\gamma\ell^{+}\ell^{-}$ final state}

In this section, we consider the $\gamma\gamma\ell^{+}\ell^{-}$ final states, which is generated by several production and decay modes

\begin{eqnarray}
\phi \gamma \rightarrow (\gamma Z) \gamma \rightarrow ( \gamma (\mu\mu) ) \gamma, & \phi Z \rightarrow (\gamma \gamma) (e e)  \\
\phi \gamma \rightarrow (\gamma Z) \gamma \rightarrow ( \gamma (ee) ) \gamma, & \phi Z \rightarrow (\gamma \gamma) (\mu \mu) 
\end{eqnarray}

Note the difference between the $( \gamma (ee) ) \gamma$ mode, where the $\phi$ is reconstructed from a photon and an electron-positron pair consistent with a $Z$ boson, and the $(\gamma \gamma) (e e)$ mode, where the $\phi$ is reconstructed from the two photons.

\begin{figure}[h!]
    \centering
    \scalebox{0.43}{\includegraphics{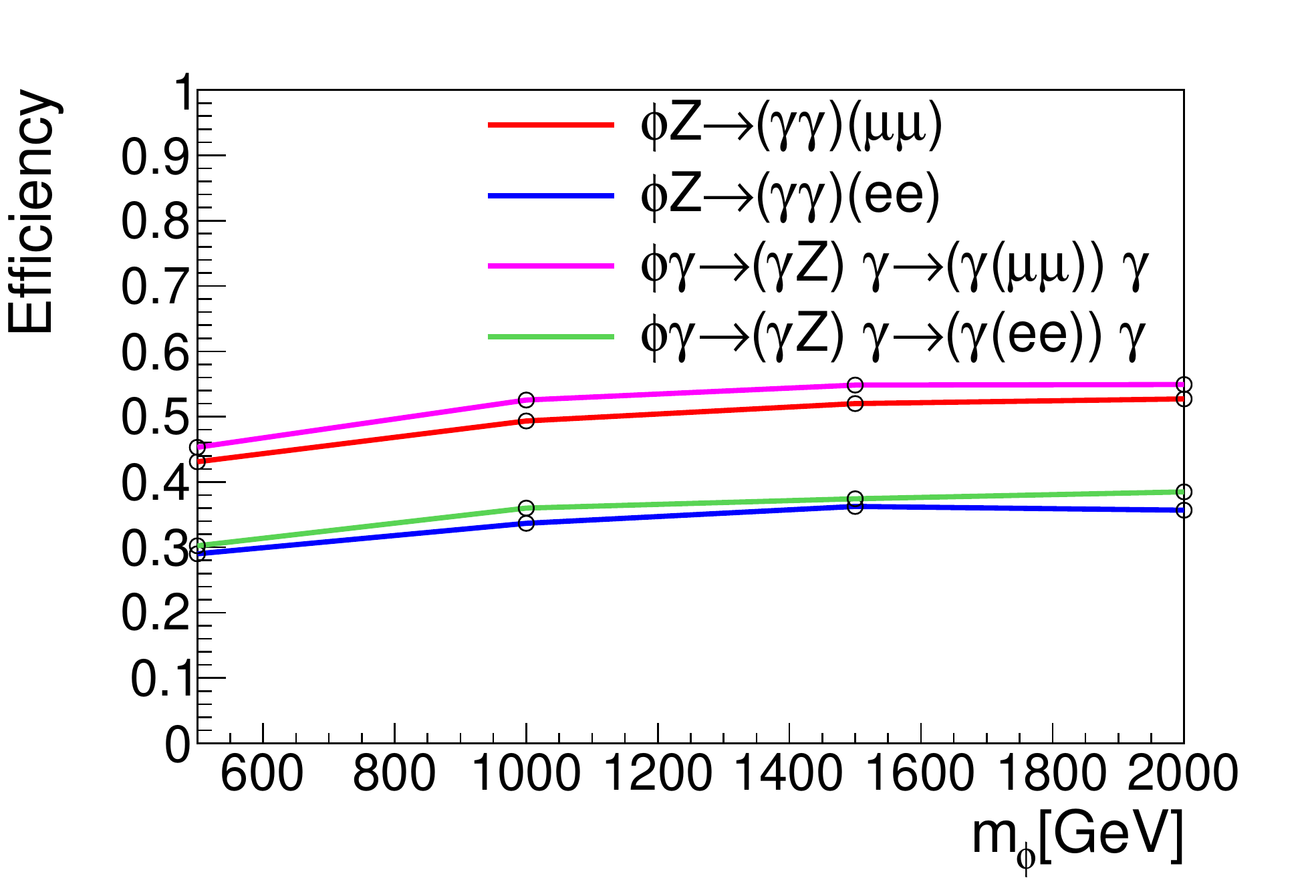}}
    \caption{ Efficiency of $\gamma\gamma\ell\ell$ selection as a function of the hypothetical $\phi$ mass for the four production and decay modes which produce a $\gamma\gamma\ell^{+}\ell^{-}$ final state; see Table~\ref{tab:modes}.}
    \label{fig:2gamma2lep_effs}
\end{figure}

The selection requires at least two photons and at least two oppositely-charged, same-flavor leptons. The major background process for this final state is SM $\gamma\gamma Z \rightarrow \gamma\gamma\ell^{+}\ell^{-}$ production. In the  $\phi \rightarrow \gamma Z$ decay chain, the $\phi$ candidate is reconstructed as the $\gamma Z$ pair with the largest $\Delta{}p_\textrm{T}$, where the $Z$ boson is reconstructed as the pair of oppositely-charged same-flavor leptons that have   invariant mass closest to $m_{Z} = 91$~GeV\footnote{\rev{No invariant mass window selection is applied, as this requirement produces reconstructed $Z$ candidates with mass close to $m_Z$}}. In $\phi \rightarrow \gamma\gamma$ decay chain, the $\phi$ candidate is reconstructed as the $\gamma\gamma$ pair with the largest $\Delta{}p_\textrm{T}$. The efficiency of the selection versus $\phi$ mass is shown in Fig.~\ref{fig:2gamma2lep_effs}. 

\begin{figure}[h!]
    \centering
    \begin{subfigure}{0.4\textwidth}
        \includegraphics[width=\textwidth]{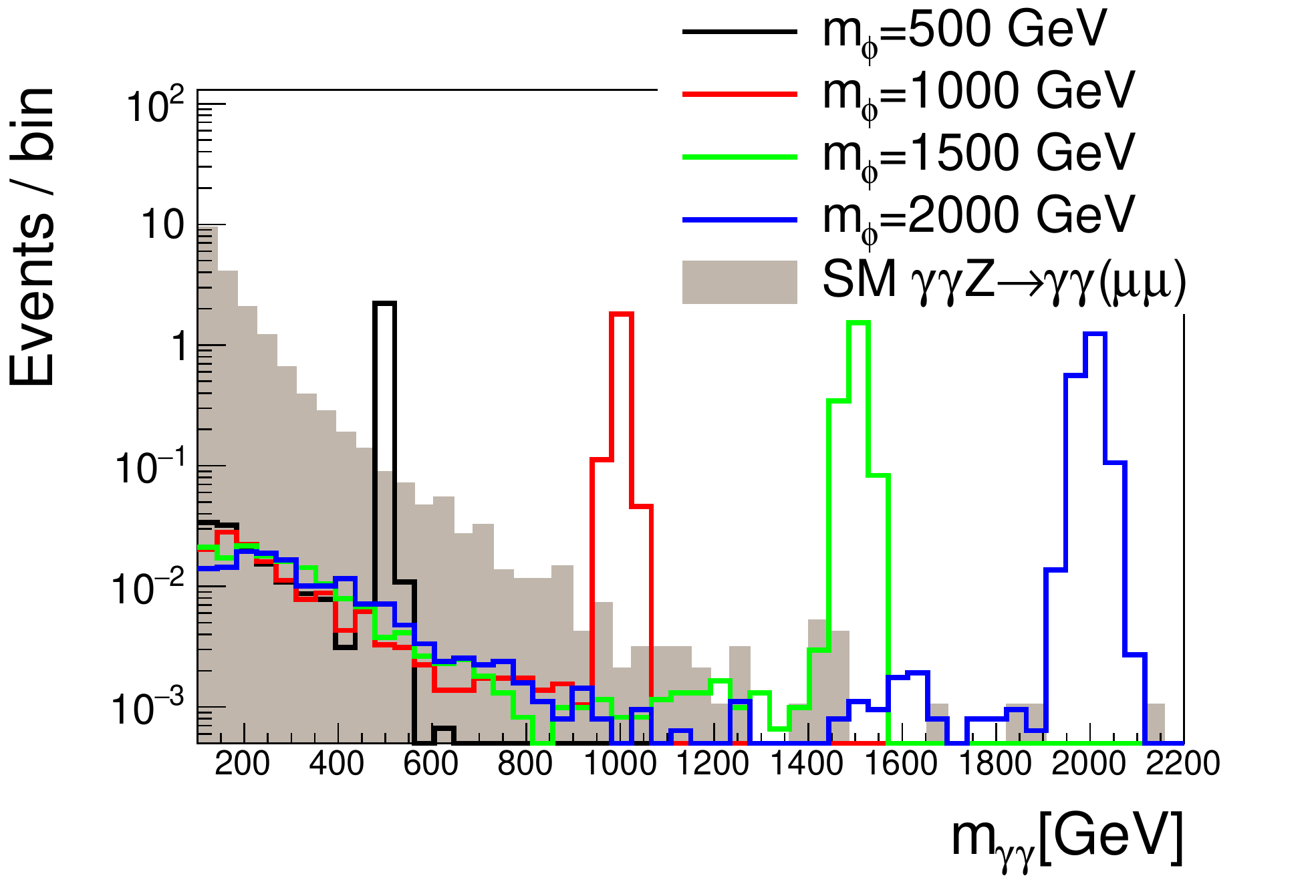}
        \caption{$\phi Z \rightarrow (\gamma \gamma) (\mu \mu)$}
        \label{fig:zphi-2m2a}
    \end{subfigure}
    \begin{subfigure}{0.4\textwidth}
        \includegraphics[width=\textwidth]{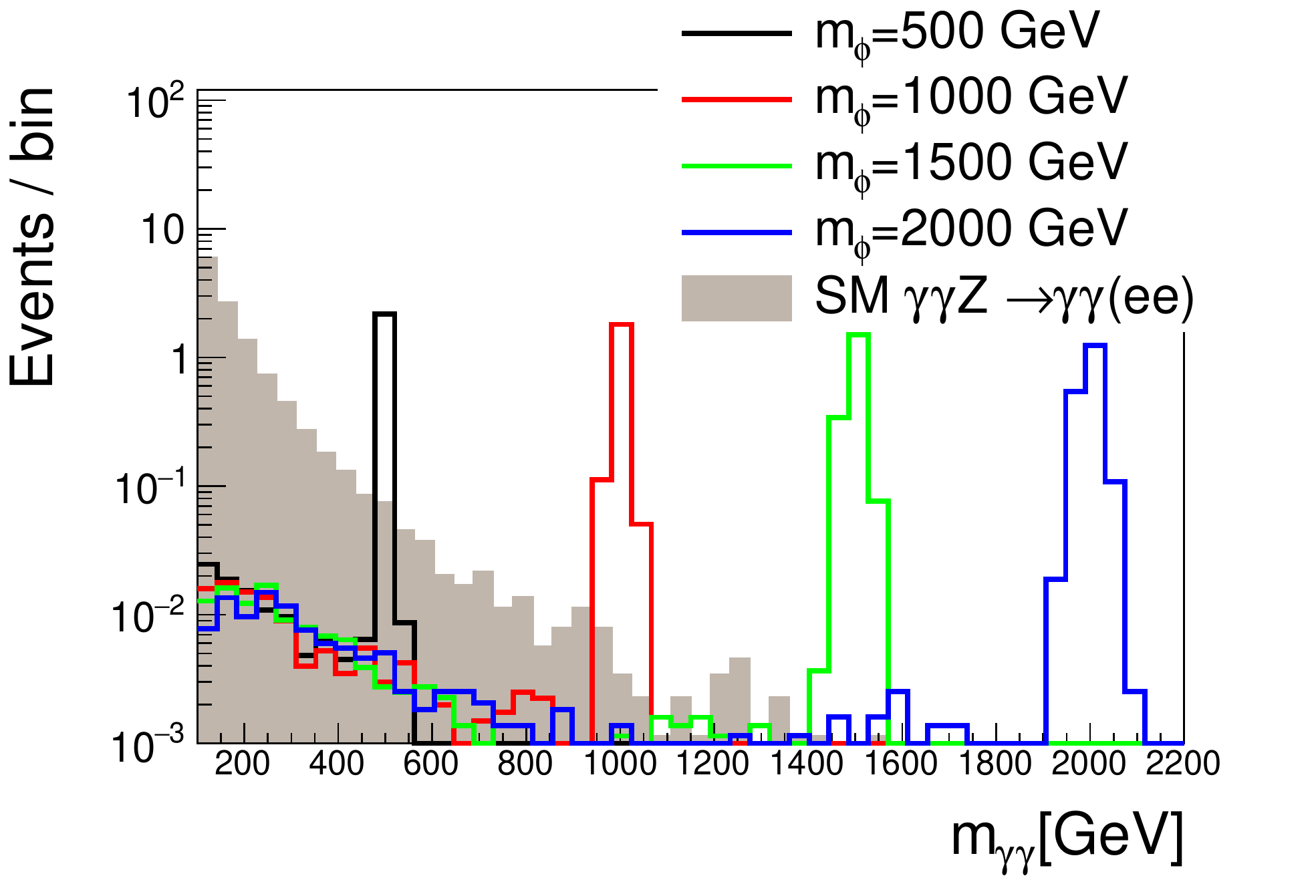}
        \caption{$\phi Z \rightarrow (\gamma \gamma) (e e)$}
        \label{fig:zphi-2e2a}
    \end{subfigure}
    \begin{subfigure}{0.4\textwidth}
        \includegraphics[width=\textwidth]{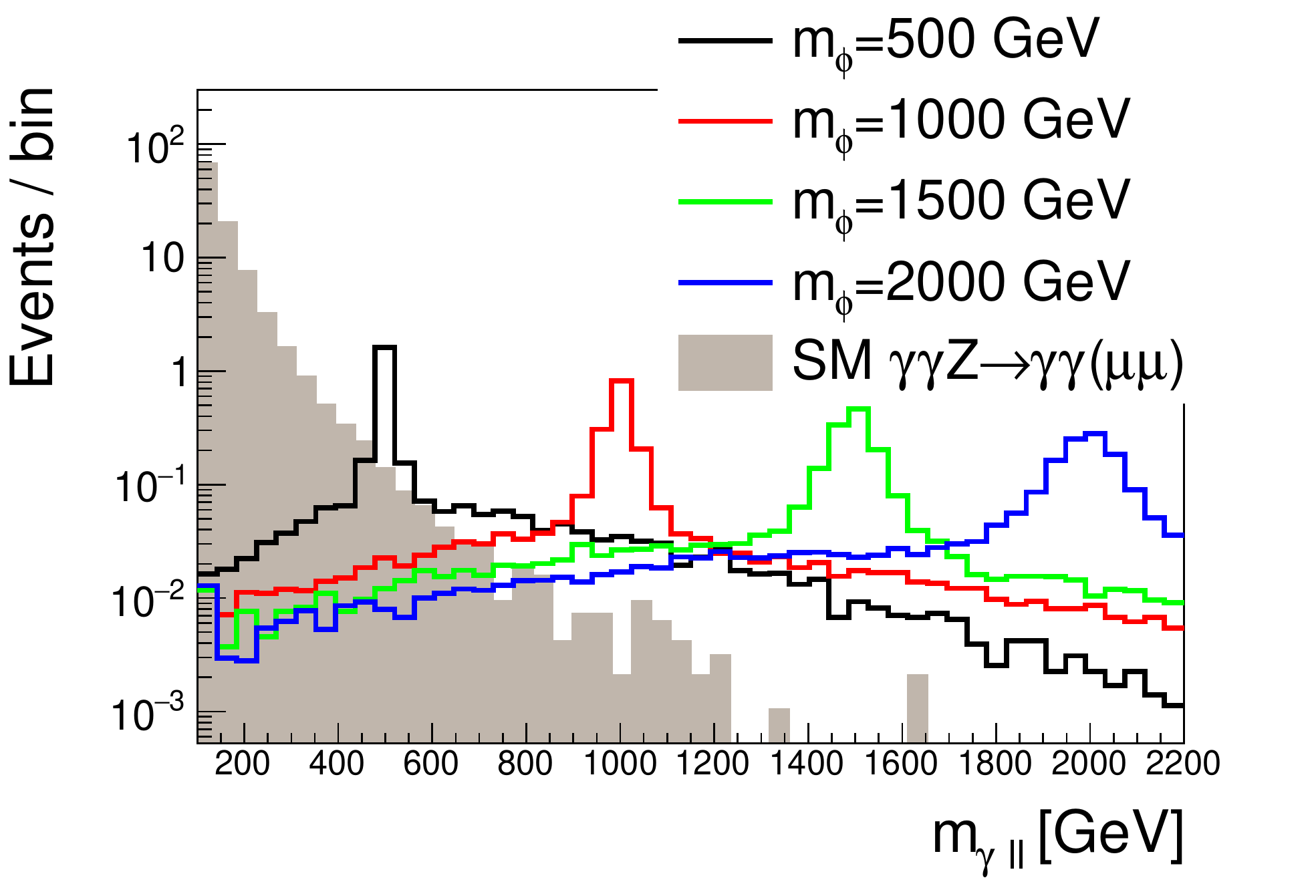}
        \caption{$\phi \gamma \rightarrow (\gamma Z) \gamma \rightarrow ( \gamma (\mu\mu) ) \gamma$}
        \label{fig:aphi-2m2a}
    \end{subfigure}
    \begin{subfigure}{0.4\textwidth}
        \includegraphics[width=\textwidth]{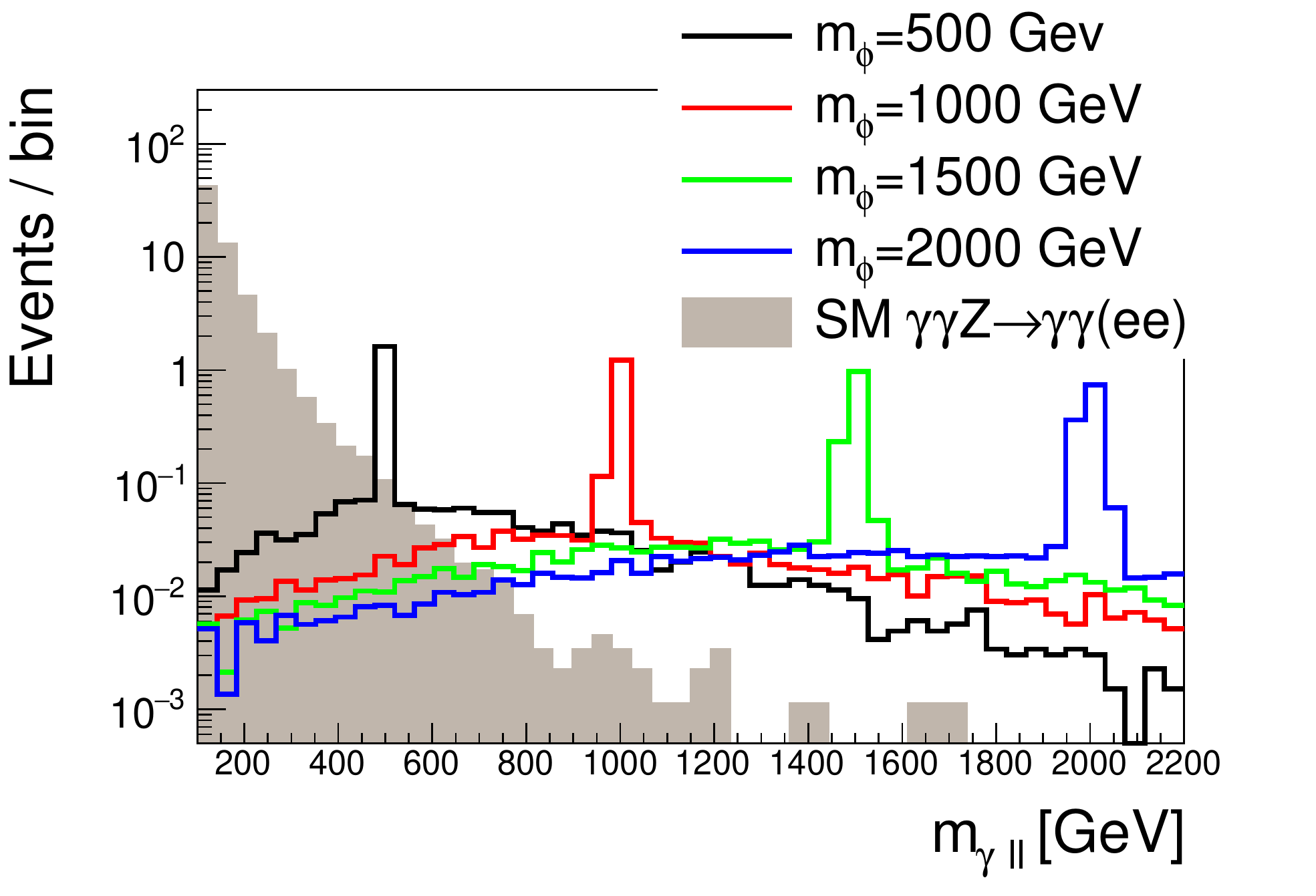}
        \caption{$\phi \gamma \rightarrow (\gamma Z) \gamma \rightarrow ( \gamma (ee) ) \gamma$ }
        \label{fig:aphi-2e2a}
    \end{subfigure}
    \caption{Distributions of reconstructed $m_\phi$ in simulated signal and background samples in $\gamma\gamma\ell^{+}\ell^{-}$ final states, normalized to integrated luminosity of 100 fb$^{-1}$. \rev{The signal cross section is calculated at leading order assuming the default values of the coupling constants; see text for details.}}
    \label{fig:2gamma2lep_plots}
\end{figure}

The distributions of reconstructed $m_\phi$ for signal and background samples are shown in in Figure~\ref{fig:2gamma2lep_plots}.  The SM background falls smoothly and rapidly, similarly to the $\gamma\gamma\gamma$ spectrum. The signal in the $\phi Z$ modes where the $\phi$ decays via $\gamma\gamma$ is very narrowly peaked due to the excellent photon energy resolutions, with broader tails in cases where a low-energy photon from initial- or final-state radiation is selected.  The signal in the $\phi \gamma$ modes where the $\phi$ decays via $\gamma Z$ is somewhat broader in the case where $Z\rightarrow \mu\mu$, due to the degraded muon  resolution at high momentum, and has a broader shoulder due to cases where the incorrect $\gamma$ is paired with the $Z$ boson. 

\subsection{$\gamma\ell^{+}\ell^{-}\ell^{+}\ell^{-}$ final state}

In this section, we consider the $\gamma\ell^{+}\ell^{-}\ell^{+}\ell^{-}$ final state, which is generated by several production and decay modes

\begin{eqnarray}
 \phi \gamma \rightarrow (ZZ) \gamma \rightarrow ( (ee) (ee) ) \gamma, & \phi Z \rightarrow (Z \gamma) Z \rightarrow ( (ee) \gamma ) (ee) \\
\phi \gamma \rightarrow (ZZ) \gamma \rightarrow ( (ee) (\mu\mu) ) \gamma, &  \phi Z \rightarrow (Z \gamma) Z \rightarrow ( (ee) \gamma ) (\mu\mu)\\
\phi \gamma \rightarrow (ZZ) \gamma \rightarrow ( (\mu\mu) (\mu\mu) ) \gamma, & \phi Z \rightarrow (Z \gamma) Z \rightarrow ( (\mu\mu) \gamma ) (ee) \\
& \phi Z \rightarrow (Z \gamma) Z \rightarrow ( (\mu\mu) \gamma ) (\mu\mu) 
\end{eqnarray}

\noindent
in which the $\phi$ resonance is reconstructed either from a pair of $Z$ bosons or a photon and a $Z$.

The selection requires at least one photon and at least two pairs of oppositely-charged same-flavor leptons. The major background process for this final state is SM $\gamma ZZ\rightarrow~\gamma~\ell^{+}\ell^{-}\ell^{+}\ell^{-}$ production. In the decay chain where $\phi\rightarrow~ZZ\rightarrow\ell^{+}\ell^{-}$, the $\phi$ candidate is reconstructed with the pair of $Z$ bosons with the largest $\Delta p_\textrm{T}$, where each $Z$ boson is reconstructed with the two pairs of oppositely-charged same-flavor leptons that have the closest invariant mass to $m_{Z} = 91$~GeV. In the decay chain where $\phi\rightarrow\gamma Z$, the $\phi$ candidate is reconstructed with the $\gamma Z$ pair with the largest $\Delta{}p_\textrm{T}$, where the $Z$ boson is reconstructed with the pair of oppositely-charged same-flavor leptons that have the closest invariant mass to $m_{Z} = 91$~GeV. 

\begin{figure}[h!]
    \centering
    \scalebox{0.43}{\includegraphics{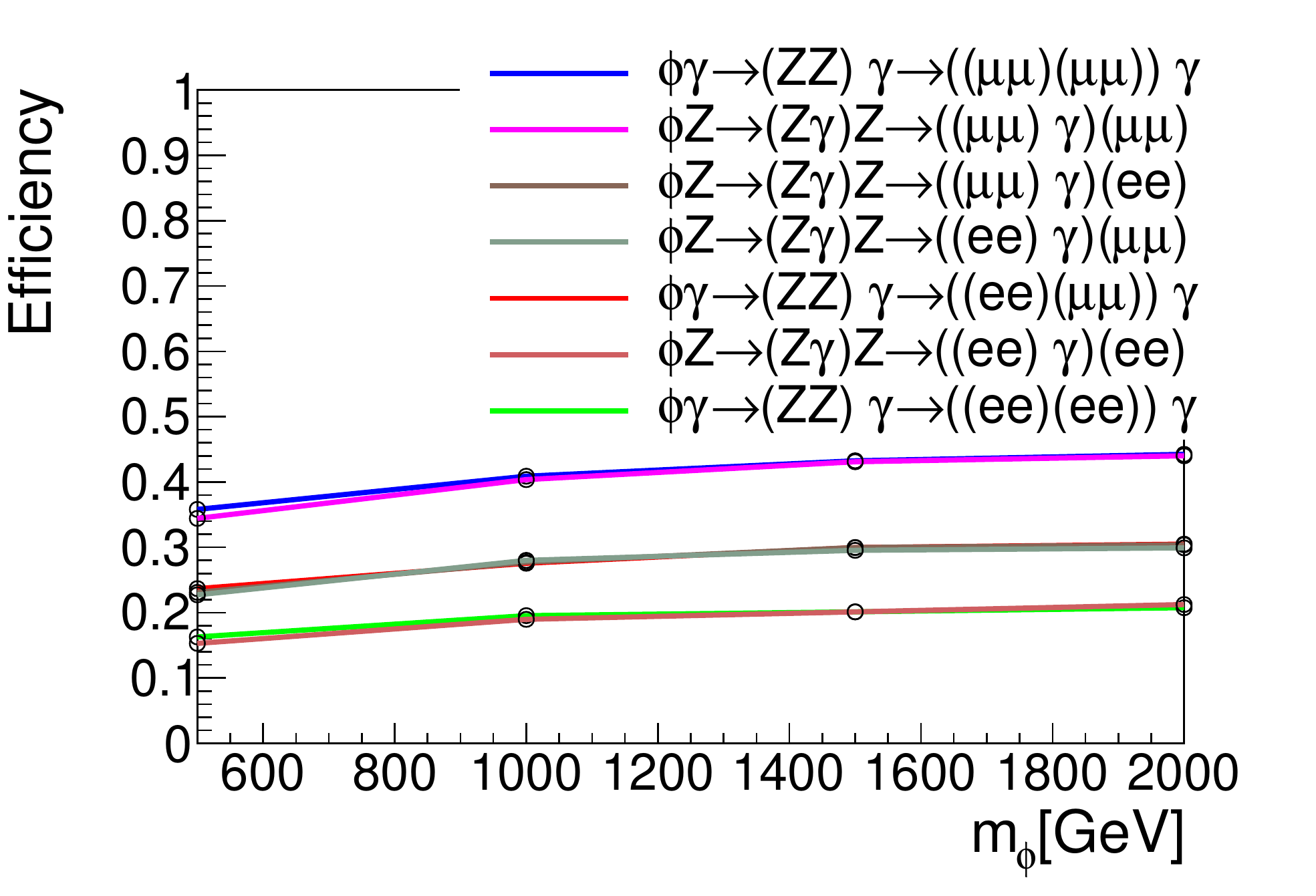}}
    \caption{ Efficiency of $\gamma 4\ell$ selection as a function of the hypothetical $\phi$ mass for the four production and decay modes which produce a $\gamma\ell^{+}\ell^{-}\ell^{+}\ell^{-}$ final state; see Table~\ref{tab:modes}.}
    \label{fig:1gamma4lep_effs}
\end{figure}

\begin{figure}[h!]
    \centering
    \begin{subfigure}{0.3\textwidth}
        \scalebox{0.25}{\includegraphics{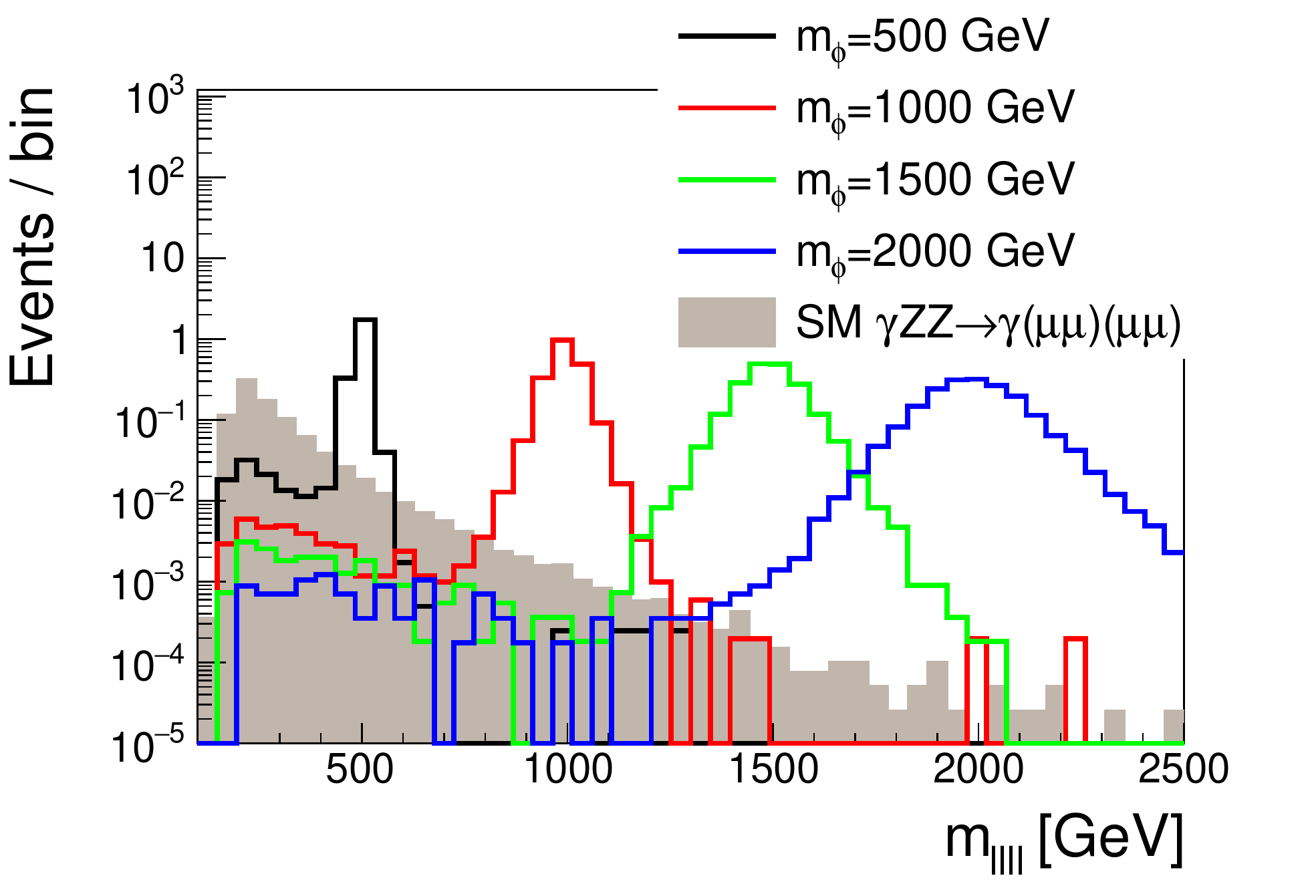}}
        \caption{$\phi \gamma \rightarrow (ZZ) \gamma \rightarrow ( (\mu\mu) (\mu\mu) ) \gamma$}
        \label{fig:aphi-1a4m}
    \end{subfigure}
    \begin{subfigure}{0.3\textwidth}
        \scalebox{0.25}{\includegraphics{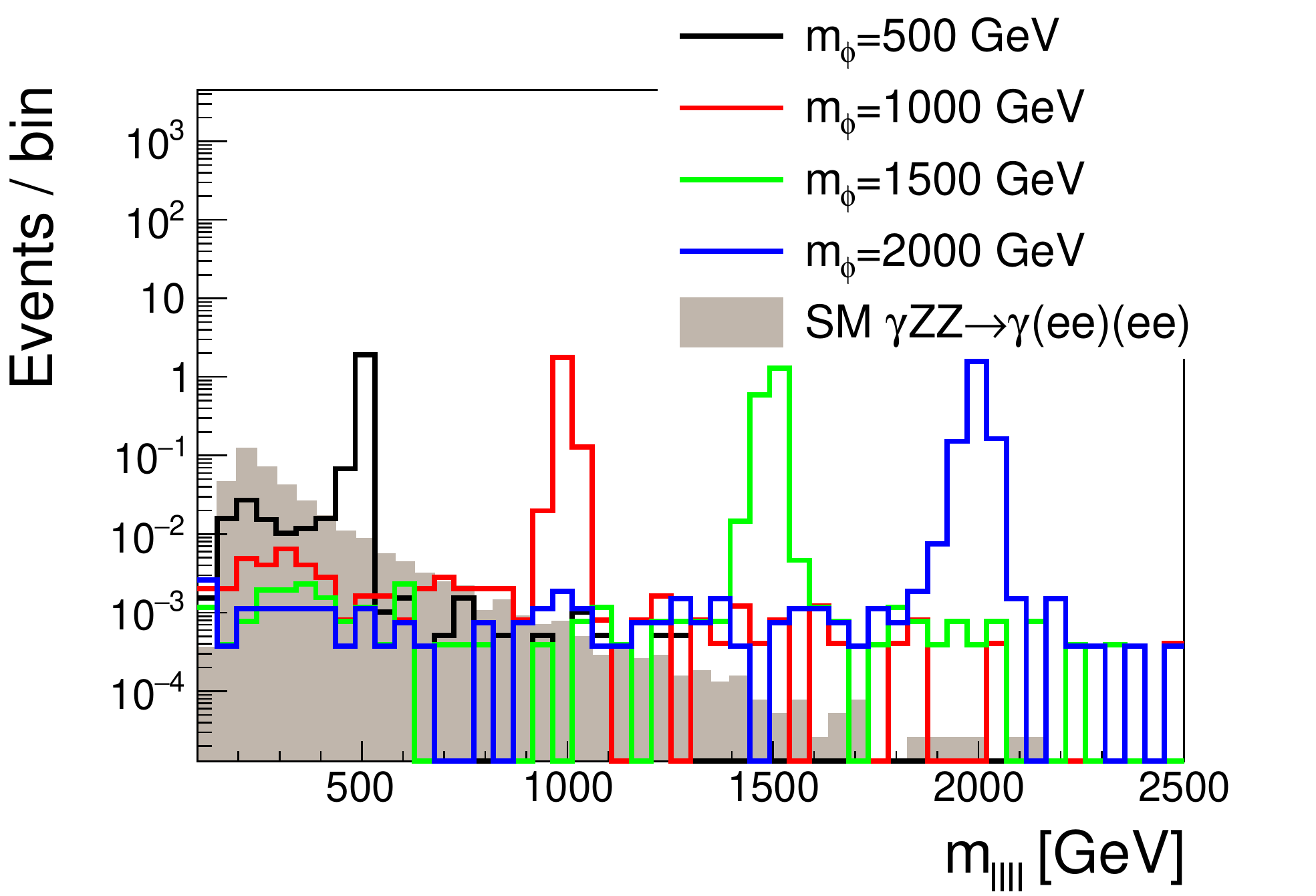}}
        \caption{$\phi \gamma \rightarrow (ZZ) \gamma \rightarrow ( (ee) (ee) ) \gamma$}
        \label{fig:aphi-1a4e}
    \end{subfigure}
    \begin{subfigure}{0.3\textwidth}
        \scalebox{0.25}{\includegraphics{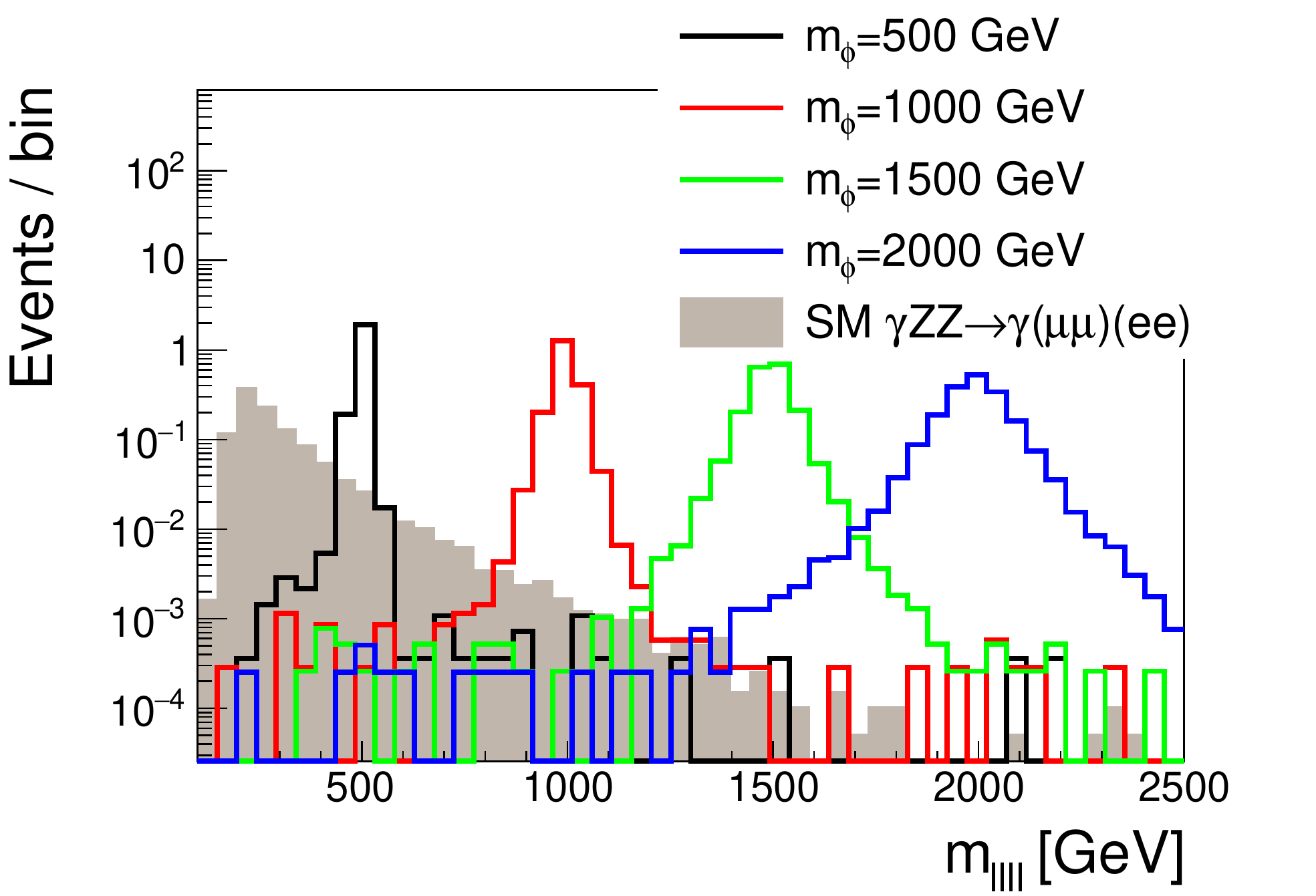}}
        \caption{$\phi \gamma \rightarrow (ZZ) \gamma \rightarrow ( (ee) (\mu\mu) ) \gamma$}
        \label{fig:aphi-1a2m2e}
    \end{subfigure}
    \caption{Distributions of reconstructed $m_\phi$ in simulated signal and background samples in $\gamma\ell^{+}\ell^{-}\ell^{+}\ell^{-}$ final states in which the $\phi$ is reconstructed from a pair of $Z$ bosons, normalized to integrated luminosity of 100 fb$^{-1}$. \rev{The signal cross section is calculated at leading order assuming the default values of the coupling constants; see text for details.}}
    \label{fig:1gamma4lep_plotsA}
\end{figure}

\begin{figure}[h!]
    \centering
    \begin{subfigure}{0.4\textwidth}
        \scalebox{0.4}{\includegraphics{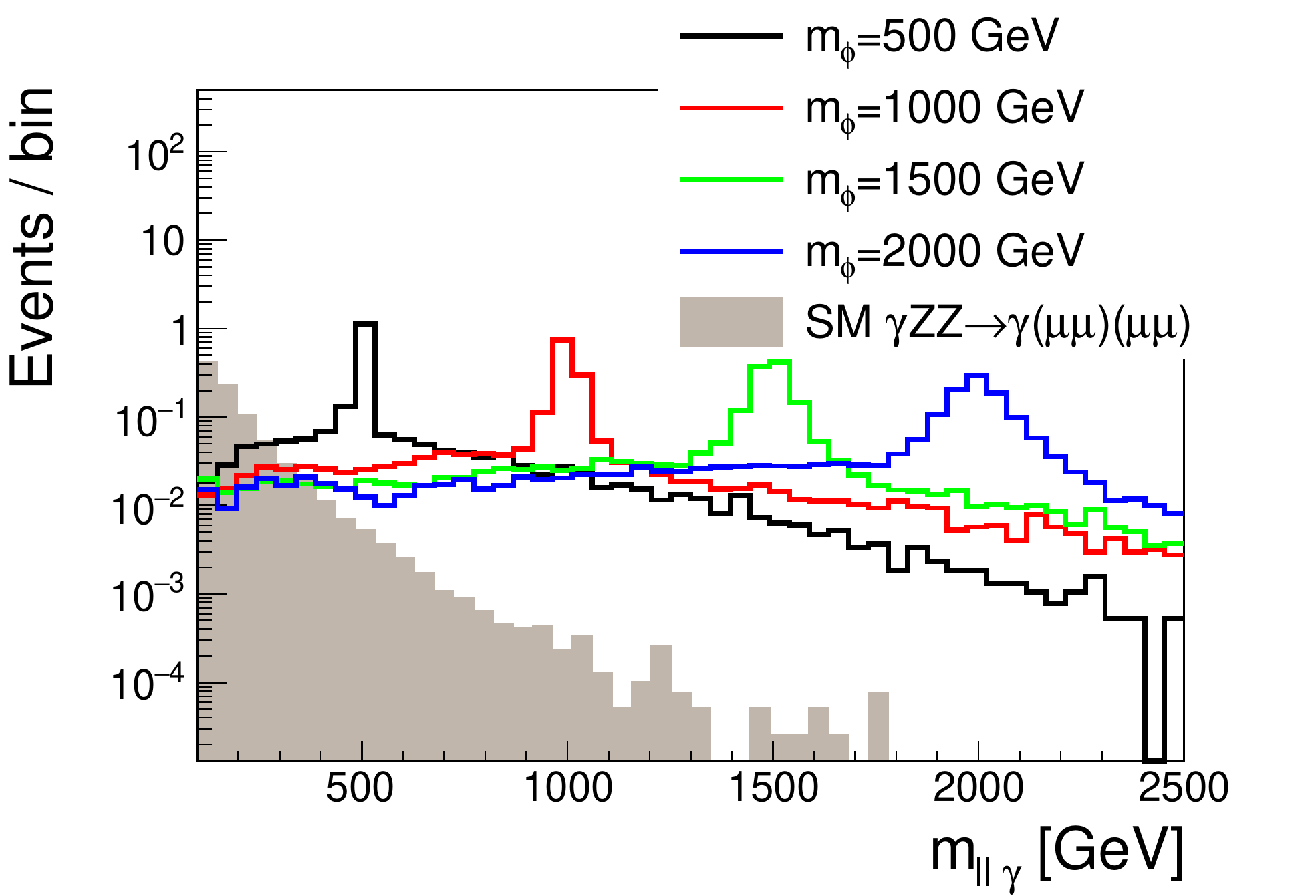}}
        \caption{$\phi Z \rightarrow (Z \gamma) Z \rightarrow ( (\mu\mu) \gamma ) (\mu\mu)$}
        \label{fig:zphi-4m1a}
    \end{subfigure}
    \begin{subfigure}{0.4\textwidth}
        \scalebox{0.4}{\includegraphics{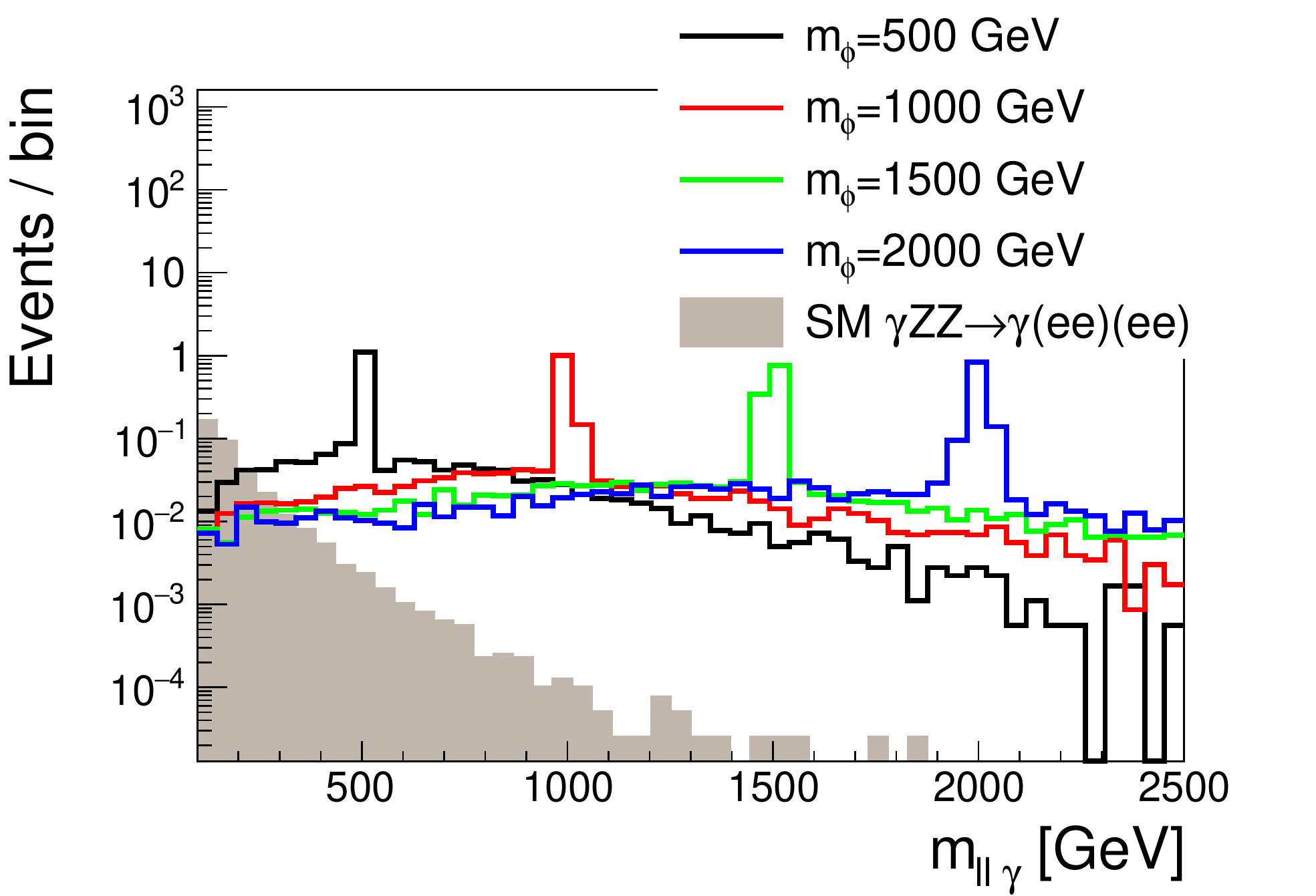}}
        \caption{$\phi Z \rightarrow (Z \gamma) Z \rightarrow ( (ee) \gamma ) (ee)$}
        \label{fig:aphi-4e1a}
    \end{subfigure}
    \begin{subfigure}{0.4\textwidth}
        \scalebox{0.4}{\includegraphics{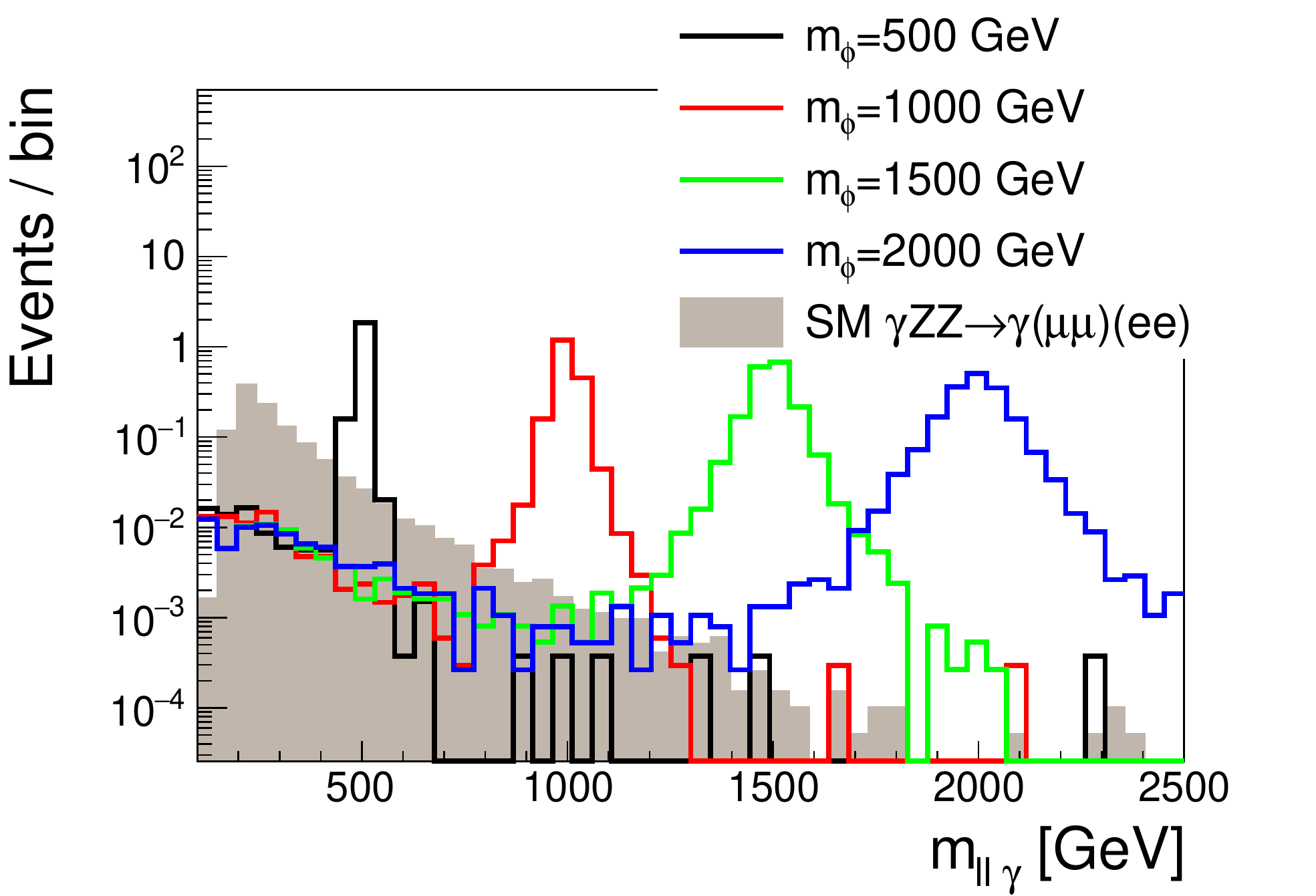}}
        \caption{$\phi Z \rightarrow (Z \gamma) Z \rightarrow ( (\mu\mu) \gamma ) (ee)$}
        \label{fig:aphi-2e2m1a}
    \end{subfigure}
    \begin{subfigure}{0.4\textwidth}
        \scalebox{0.4}{\includegraphics{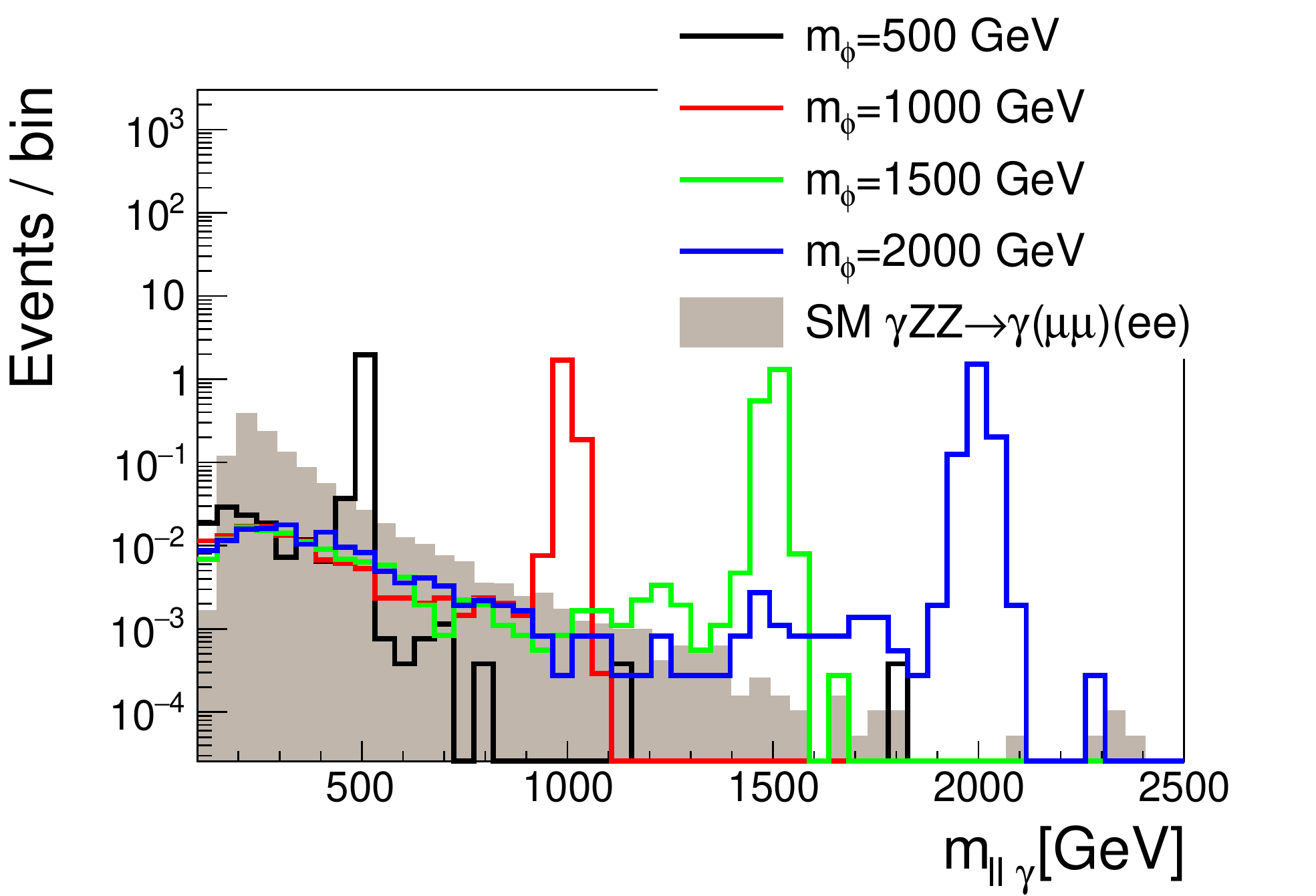}}
        \caption{$\phi Z \rightarrow (Z \gamma) Z \rightarrow ( (ee) \gamma ) (\mu\mu)$}
        \label{fig:aphi-2m2a1a}
    \end{subfigure}
    \caption{Distributions of reconstructed $m_\phi$ in simulated signal and background samples in $\gamma\ell^{+}\ell^{-}\ell^{+}\ell^{-}$ final states in which the $\phi$ is reconstructed from a $\gamma$ and $Z$ bosons, normalized to integrated luminosity of 100 fb$^{-1}$. \rev{The signal cross section is calculated at leading order assuming the default values of the coupling constants; see text for details.}}
    \label{fig:1gamma4lep_plotsB}
\end{figure}

The efficiency of the selection versus $\phi$ mass is shown in Fig.~\ref{fig:1gamma4lep_effs}. The distributions of reconstructed $m_\phi$ for signal and background samples are shown in in Figures~\ref{fig:1gamma4lep_plotsA} and \ref{fig:1gamma4lep_plotsB}.  The SM background fall smoothly as expected in each case. The varying widths of the reconstructed peaks reflect the greater (weaker) resolution for electrons (muons) at high momentum. As expected, shoulders due to mistaken assignments occur more often in cases with greater possible degeneracy, such as states with four electrons and muons, and less often in cases where lepton-flavor can distinguish, such as those with two electons and two muons.

\subsection{$\gamma\ell^{+}\ell^{-}J$ final state}

In this section, we consider the $\phi \gamma \rightarrow (ZZ) \gamma \rightarrow ( J (ee) ) \gamma$ and $\phi \gamma \rightarrow (ZZ) \gamma \rightarrow ( J (\mu\mu) ) \gamma$  modes, which lead to the $\gamma\ell^{+}\ell^{-}J$ final state.

The selection requires at least one photon, at least two oppositely-charged same-flavor leptons, and one large-radius jet. The major background process for this final state is SM $\gamma Z j j\rightarrow\gamma\ell^{+}\ell^{-}j j$ production, where the pair of jets due to QCD radiation happen to be reconstructed as a single massive large-radius jet $J$; the corresponding process with  SM $\gamma Z Z\rightarrow\gamma\ell^{+}\ell^{-}j j$ gives a much smaller predicted contribution due to the additional weak boson, despite the large-radius jet corresponding to a true massive boson decay. 

\begin{figure}[h!]
    \centering
    \scalebox{0.43}{\includegraphics{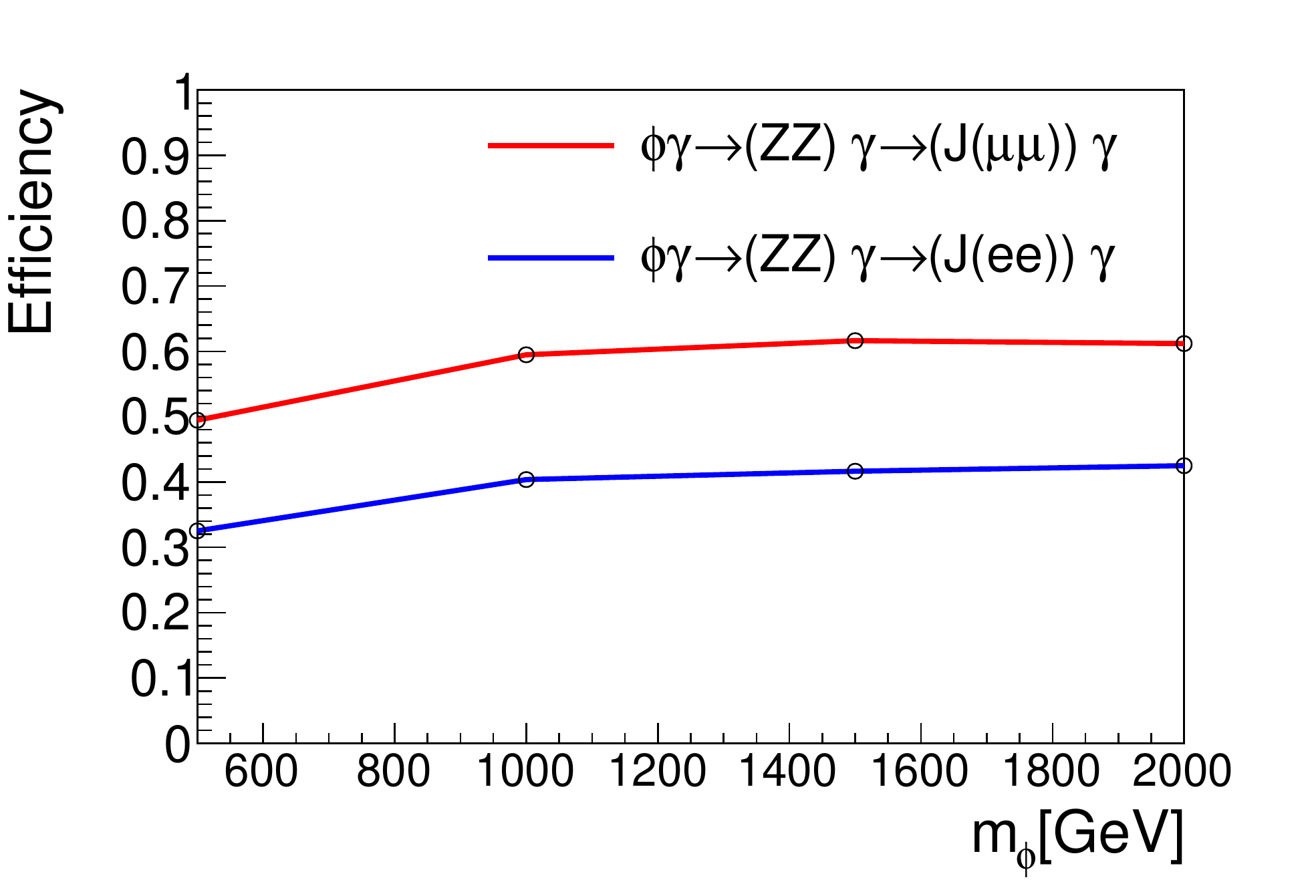}}
    \caption{Efficiencies of $\phi$ selection as functions of the $\phi$ mass for the two production and decay modes with produce  $\gamma\ell^{+}\ell^{-}J$ final states.}
    \label{fig:1gamma2lep1J_effs}
\end{figure}

The $\phi$ candidate is reconstructed from the leptonic and hadronic $Z$ boson candidates which have the largest $\Delta p_\textrm{T}$. The leptonic $Z$ boson is reconstructed from the pair of oppositely-charged same-flavor leptons that have the closest invariant mass to $m_{Z}$, and the hadronic $Z$ boson candidate is the large-radius jet with invariant mass closest to $m_{Z}$. The efficiency of the selection versus $\phi$ mass is shown in Fig.~\ref{fig:1gamma2lep1J_effs}. 

\begin{figure}[h!]
    \centering
    \begin{subfigure}{0.45\textwidth}
        \scalebox{0.4}{\includegraphics{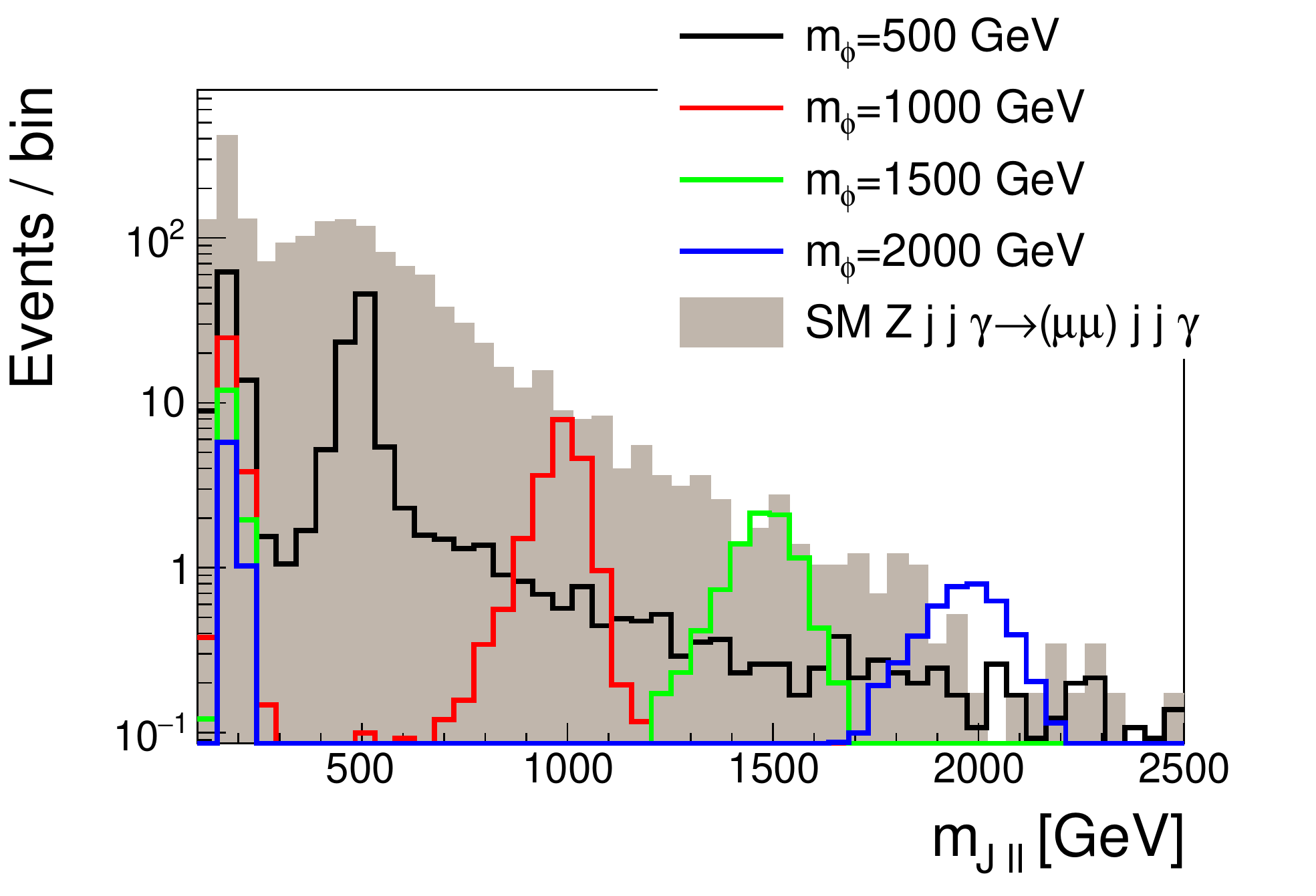}}
        \caption{$\phi \gamma \rightarrow (ZZ) \gamma \rightarrow ( J (\mu\mu) ) \gamma$}
        \label{fig:aphi-1a2m1J}
    \end{subfigure}
    \begin{subfigure}{0.45\textwidth}
        \scalebox{0.4}{\includegraphics{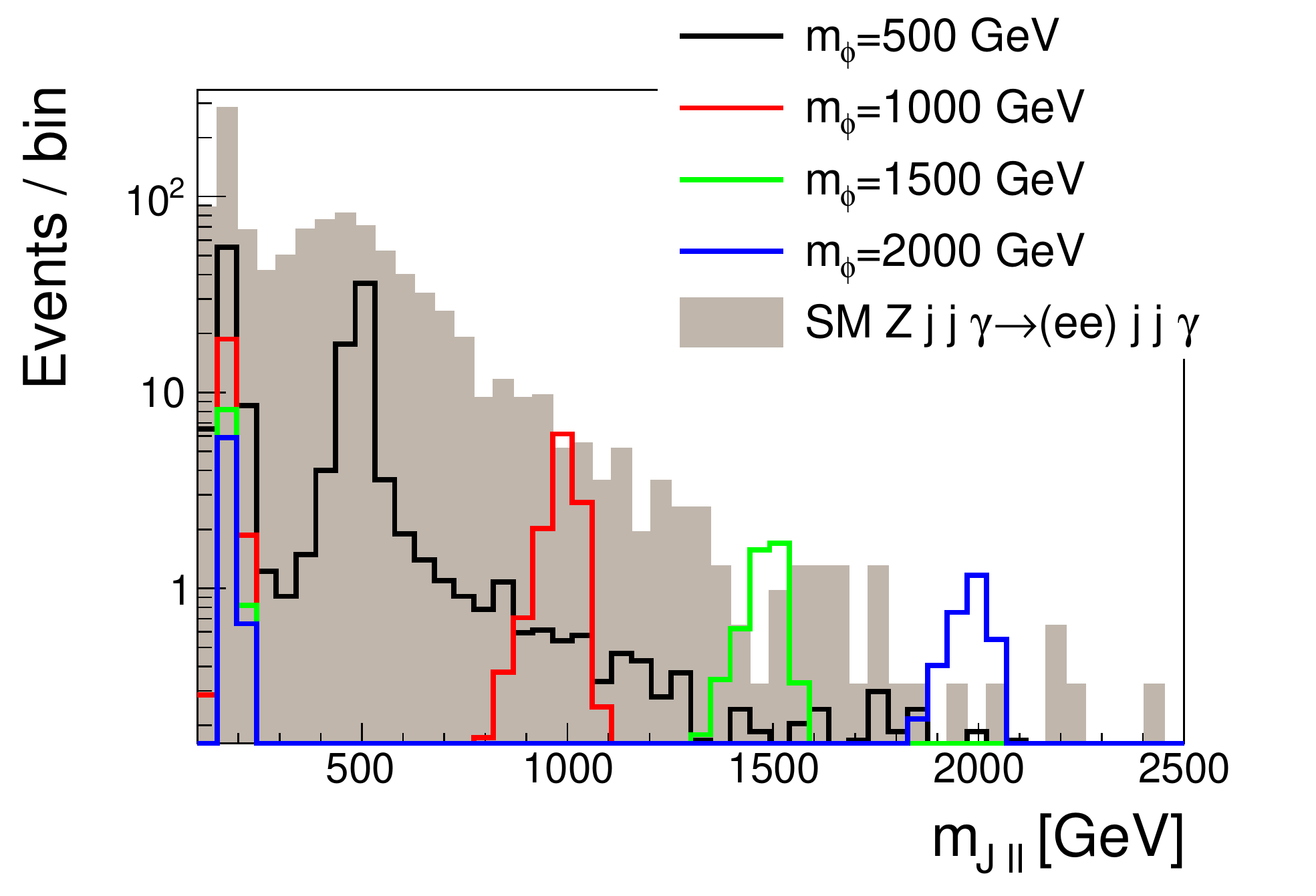}}
        \caption{$\phi \gamma \rightarrow (ZZ) \gamma \rightarrow ( J (ee) ) \gamma$ }
        \label{fig:aphi-1a2m1J}
    \end{subfigure}
    \caption{Distributions of reconstructed $m_\phi$ in simulated signal and background samples in $\gamma\ell^{+}\ell^{-}J$ final states, normalized to integrated luminosity of 100 fb$^{-1}$. \rev{The signal cross section is calculated at leading order assuming the default values of the coupling constants; see text for details.}}
    \label{fig:1gamma2lep1J_plots}
\end{figure}

The distributions of reconstructed $m_\phi$ for signal and background samples are shown in in Figures~\ref{fig:1gamma2lep1J_plots}.  The spike at small values of $m_{J\ell\ell}$ are due to cases where the large-radius jet fails to capture the hadronic decay of the $Z$ boson, leading to a mass close to $m_{Z}$ from the leptonically-decaying $Z$ boson.

\subsection{Statistical Analysis}

\begin{figure}
    \centering
    \scalebox{0.43}{\includegraphics{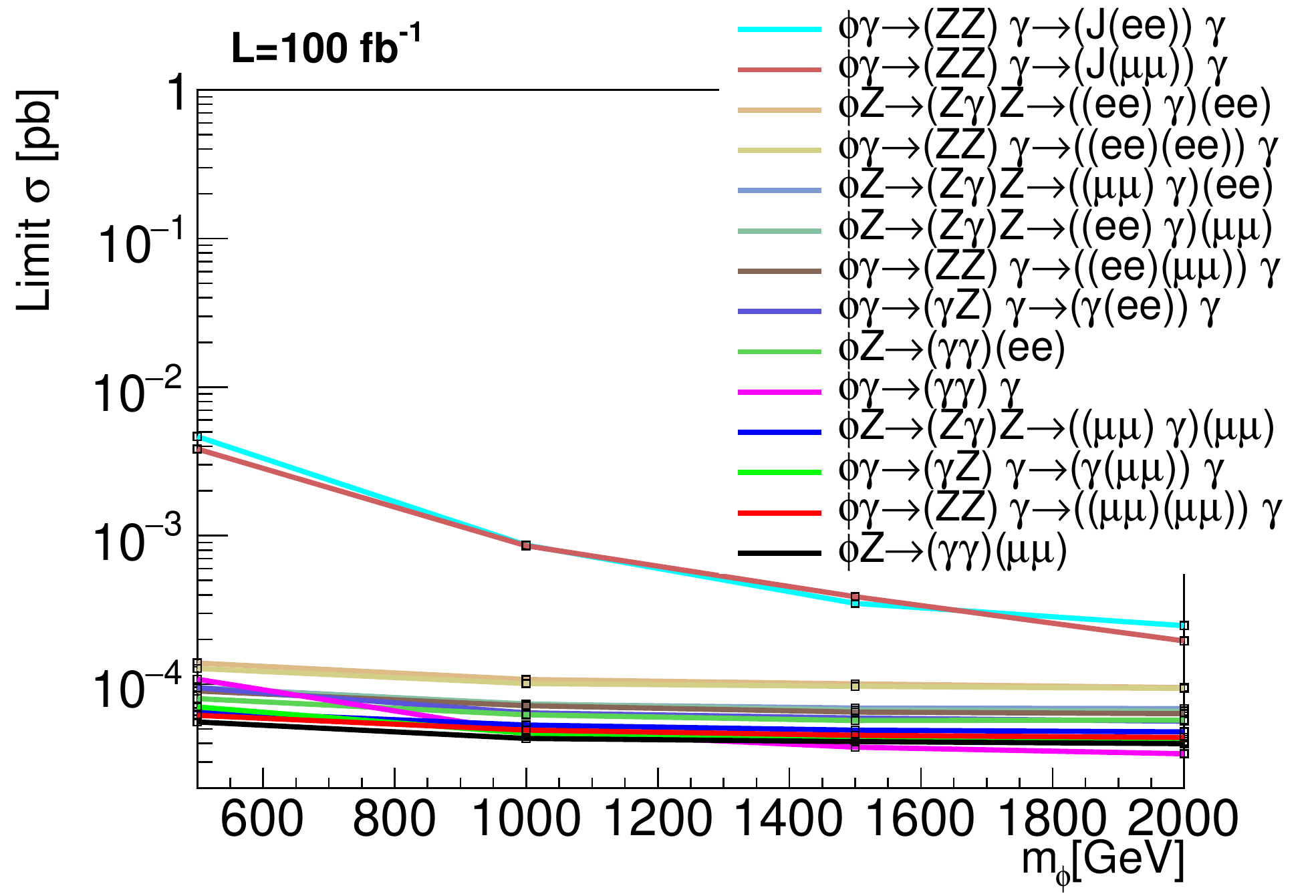}}
    \caption{Summary of expected upper limits at 95\% CL on the production cross-sections as functions of the $\phi$ mass in integrated luminosity of 100 fb$^{-1}$ for the fourteen production and decay modes which produce five unique final states; see Table~\ref{tab:modes}.}
    \label{fig:xs_limit_summary}
\end{figure}

Limits are calculated at 95\% CL using a profile likelihood ratio~\cite{Cowan:2010js}  with the CLs technique~\cite{Junk:1999kv,Read:2002hq} with RooStats~\cite{Moneta:2010pm} for a binned distribution in the reconstructed mass of the hypothetical $\phi$ boson, where bins without simulated background events have been merged into adjacent bins. A graph of these limits as functions of the $\phi$ mass are shown in Figure~\ref{fig:xs_limit_summary}, and selected data points are shown in Table \ref{tab:limits}. The background is assumed to have a 50\% relative systematic uncertainty.  

\begin{table}[]
    \centering
    \begin{tabular}{c|c|c}
         \hline
                  \hline
         Mode & \multicolumn{2}{c}{95\% CL Expected Upper Limit [fb]}\\
         & $m_{\phi{}}=500$ GeV &  $m_{\phi{}}=2000$ GeV\\
         \hline
         $\phi\gamma\rightarrow(ZZ)\gamma\rightarrow(J(ee))\gamma$ & 4.6 & 2.5$\times 10^{-1}$ \\
         \hline
         $\phi\gamma\rightarrow(ZZ)\gamma\rightarrow(J(\mu\mu))\gamma$ & 4.6 & 2.5$\times 10^{-1}$ \\
         \hline
         $\phi{}Z\rightarrow(Z\gamma)Z\rightarrow((ee)\gamma)(ee)$ & 1.4$\times 10^{-1}$ & 9.5$\times 10^{-2}$ \\
         \hline
         $\phi\gamma\rightarrow(ZZ)\gamma\rightarrow((ee)(ee))\gamma$ & 1.3$\times 10^{-1}$ & 9.4$\times 10^{-2}$ \\
         \hline
         $\phi{}Z\rightarrow(Z\gamma)Z\rightarrow((\mu\mu)\gamma)(ee)$ & 9.5$\times 10^{-2}$ & 6.9$\times 10^{-2}$ \\
         \hline
         $\phi{}Z\rightarrow(Z\gamma)Z\rightarrow((ee)\gamma)(\mu\mu)$ & 9.4$\times 10^{-2}$ & 6.6$\times 10^{-2}$ \\
         \hline
         $\phi\gamma\rightarrow(ZZ)\gamma\rightarrow((ee)(\mu\mu))\gamma$ & 9.0$\times 10^{-2}$ & 6.3$\times 10^{-2}$ \\
         \hline
         $\phi\gamma\rightarrow(\gamma{}Z)\gamma\rightarrow(\gamma(ee))\gamma$ & 9.5$\times 10^{-2}$ & 5.6$\times 10^{-2}$ \\
         \hline
         $\phi{}Z\rightarrow(\gamma\gamma)(ee)$ & 8.0$\times 10^{-2}$ & 5.7$\times 10^{-2}$ \\
         \hline
         $\phi\gamma\rightarrow(\gamma\gamma)\gamma$ & 1.1$\times 10^{-1}$ & 3.4$\times 10^{-2}$ \\
         \hline
         $\phi{}Z\rightarrow(Z\gamma)Z\rightarrow((\mu\mu)\gamma)(\mu\mu)$ & 6.5$\times 10^{-2}$ & 4.8$\times 10^{-2}$ \\
         \hline
         $\phi\gamma\rightarrow(\gamma{}Z)\gamma\rightarrow(\gamma(\mu\mu))\gamma$ & 7.0$\times 10^{-2}$ & 4.1$\times 10^{-2}$ \\
         \hline
         $\phi\gamma\rightarrow(ZZ)\gamma\rightarrow((\mu\mu)(\mu\mu))\gamma$ & 6.2$\times 10^{-2}$ & 4.4$\times 10^{-2}$ \\
         \hline
         $\phi{}Z\rightarrow(\gamma\gamma)(\mu\mu)$ & 5.6$\times 10^{-2}$ & 4.0$\times 10^{-2}$ \\
         \hline
                  \hline
    \end{tabular}
    \caption{Summary of expected upper limits at 95\% CL on production cross-sections at $m_{\phi}=500$ GeV and $m_{\phi}=2000$ GeV in integrated luminosity of 100 fb$^{-1}$ for the fourteen production and decay modes which produce five unique final states; see Fig~\ref{fig:xs_limit_summary}.}
    \label{tab:limits}
\end{table}

\clearpage

\section{Interpretation}
\label{sec:interp}

We will consider separately how the models from Section 2 are constrained by our search. In Table \ref{models}, we identify the six separate models that have  triple electroweak boson signatures. These models are listed in order of increasing dimension of scalar $SU(2)$ representation and have effective operators and coefficients as listed. 

\begin{table}[H]  
\centering
\renewcommand*{\arraystretch}{2.5}
\begin{tabular}{|c|c|c|} 
\hline
Label & Description & $\mathcal{L}$ \\
\hline 
A & singlet dim 5 & $\frac{1}{\Lambda_{XBB}} X B^{\mu \nu} B_{ \mu \nu}+ \frac{1}{\Lambda_{XWW}} X W^{\mu \nu} W_{ \mu \nu}$\\ 
\hline

B & singlet dim 7 & $\frac{1}{\Lambda_{XBW}^3}  X B_{\mu \nu} [H^{\dagger}W_{ \mu \nu} H]$ \\

\hline

C & doublet I & $\frac{1}{\Lambda_{YBB}^2} [H^{\dagger}  Y] B^{\mu \nu} B_{ \mu \nu}+ \frac{1}{\Lambda_{YWW}^2} [H^{\dagger}  Y] W^{\mu \nu} W_{ \mu \nu} $\\

\hline

D & doublet II & $\frac{1}{\Lambda_{YBW}^2}   B_{\mu \nu} [H^{\dagger}W_{ \mu \nu} Y]$ \\

\hline

E & adjoint dim 5 & $\frac{1}{\Lambda_{TWB}} T_i W_i^{\mu \nu} B_{ \mu \nu} $ \\

\hline

F & adjoint dim 7 & $\frac{1}{\Lambda_{TBB}^3} [H^{\dagger} T H] B^{\mu \nu} B_{ \mu \nu}+ \frac{1}{\Lambda_{TWW}^3} [H^{\dagger} T H] W^{\mu \nu} W_{ \mu \nu}$\\

\hline 

\end{tabular}
\caption{The six effective models considered for this investigation, organized by the scalar's $SU(2)$ representation and the dimension of the effective operators.}
\label{models}
\end{table}

Here we identify the generic field $\phi$ with the neutral component of the specific exotic scalar multiplet in each model, the singlet $X$, the neutral doublet component $Y_0$ and the neutral triplet component $T_0$, that is. $\phi \supset X, Y_0, T_0$. In the following section, we will translate the cross section limits described above into bounds on parameter space for each model. For each model, we will provide the vertex coefficient rules for scalar-boson couplings, as verified by a FeynRules \cite{Alloul:2013bka} implementation of the model.

Once electroweak symmetry is broken, the Higgs fields is replaced in our operators with its vacuum expectation value and we write the gauge fields in their mass eigenbasis. After electroweak symmetry breaking, the effective cut-off parameters control the coupling of the neutral exotic scalar $\phi_0$ to 4 pairs of mass eigenstate electroweak gauge bosons: $\gamma\gamma,\gamma Z, Z Z, $ and $WW$.  Upon inspection we find that for the models listed, there are two distinct patters of couplings between the exotic scalar $\phi$ and pairs of electroweak gauge bosons.   

In models A, C, and F, scalars couple separately to $B_{\mu \nu}B^{\mu \nu}$, $W_{\mu \nu}W^{\mu \nu}$ with two distinct effective cut off parameters; these two parameters specify the couplings of the scalar $\phi$ to all four pairs of electroweak bosons. We express the pattern of diboson coupling coefficients in \rev{ Table~\ref{pattern-0}} below.

\begin{table}[H]  
\centering
\begin{tabular}{|c|} 
\hline 
Pattern 1 \\
\hline
\\
$V_{\phi\gamma \gamma} =  \kappa\left[\frac{ c_w^2} {\Lambda_{\text{1}}^n}+\frac{ s_w^2 }{\Lambda_{\text{2}}^n} \right]$
\\
\\
$V_{\phi WW} = \kappa \left[\frac{1}{\Lambda_{\text{2}}^n}  \right]$ 
\\
\\
$V_{\phi \gamma Z} = \kappa\left[-\frac{c_w s_w}{\Lambda_{\text{1}}^n} +\frac{ c_w s_w}{\Lambda_{\text{2}}^n}\right] $ 
\\
\\
$V_{\phi ZZ} = \kappa \left[\frac{ c_w^2}{\Lambda_{\text{2}}^n} +\frac{ s_w^2}{\Lambda_{\text{1}}^n} \right]$
\\
\\
\hline 

\end{tabular} 
\quad
\begin{tabular}{|c|}
\hline
     dimension 5 scalar \\
     \hline
     \\
     $n=1$ $\kappa=1$\\
     \\
     \hline
     weak doublet I\\
     \hline
     \\
     $n=2$ $\kappa=v_h$\\
     \\
     \hline
    dimension 7 adjoint\\
    \hline
    \\
     $n=3$ $\kappa=-v_h^2/\sqrt{32}$ \\
     \\
     \hline
     
\end{tabular}
\caption{The effective couplings of the models A, C, and F, in terms of effective cutoffs $\Lambda_1$ and $\Lambda_2$ (left) and values for coefficients $\kappa$ and powers n for benchmark models (right).}
\label{pattern-0}
\end{table}

\begin{figure}[h!]
    \centering
    \includegraphics[width=0.5\linewidth]{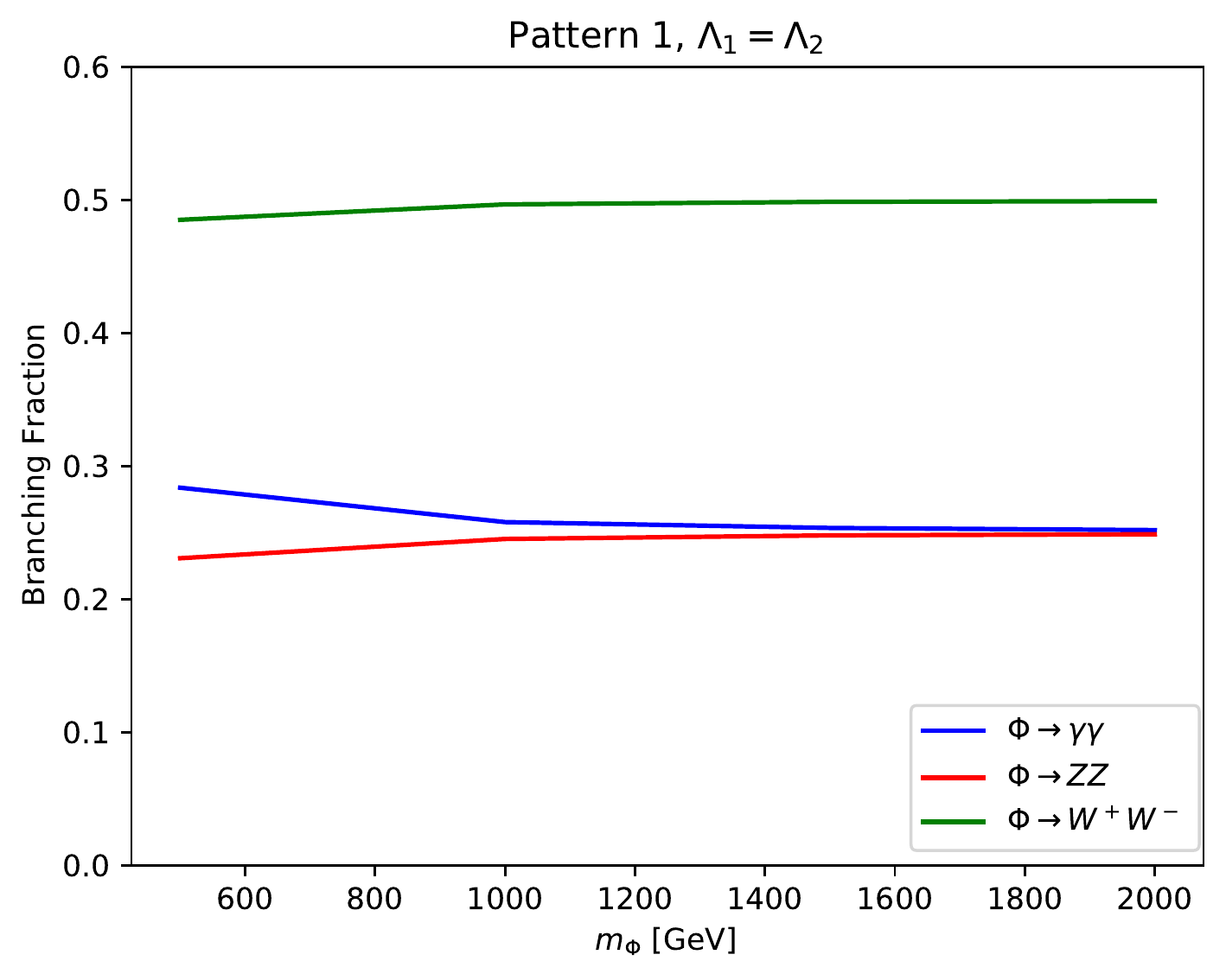}
    \includegraphics[width=0.5\linewidth]{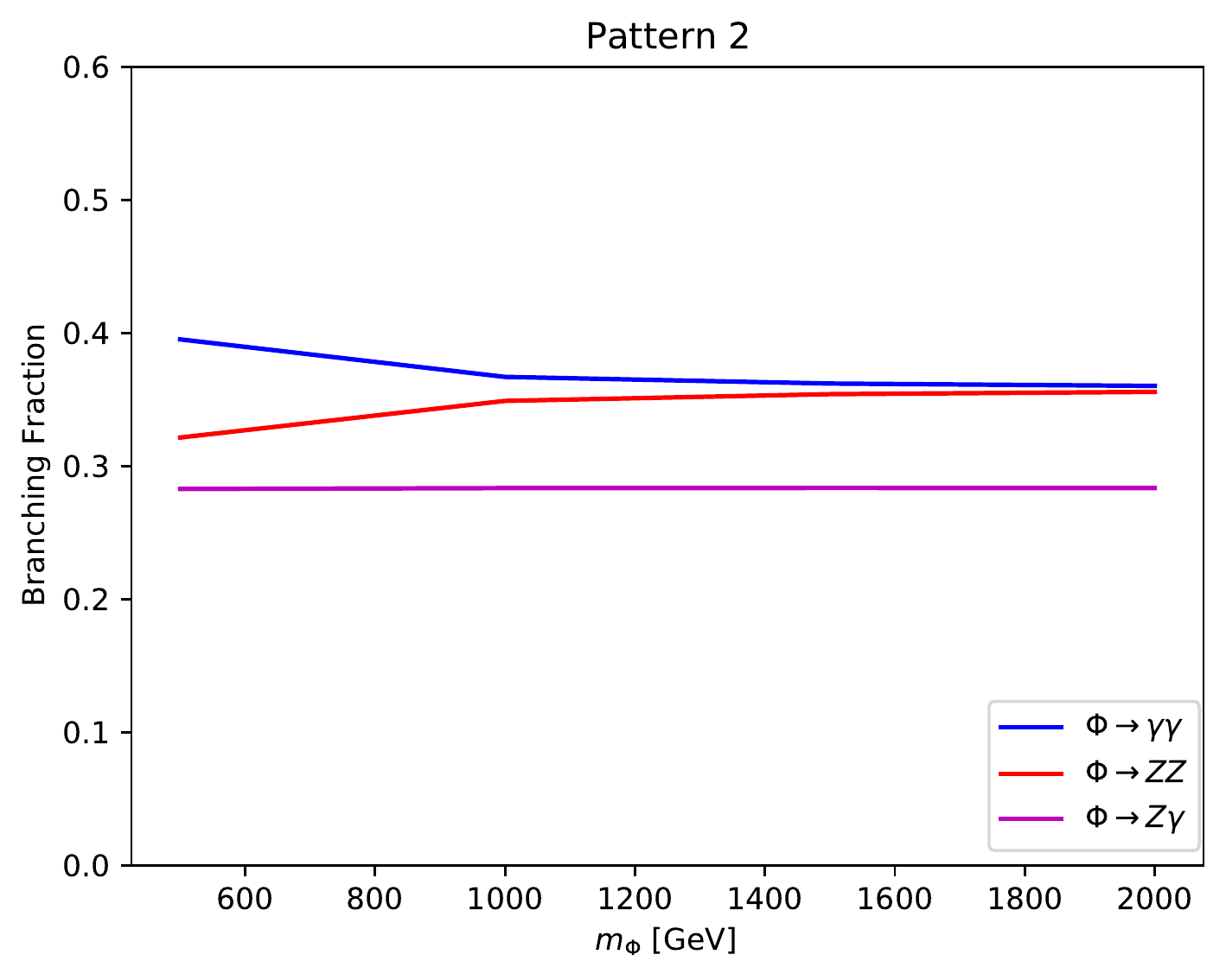}
    \caption{Branching fractions of the scalar, assuming (top) pattern 1 (models A, C, and F) and (bottom) pattern 2 (models B, D, and E).}
    \label{bf-plots}
\end{figure}

Here the scales $\Lambda_1$ and $\Lambda_2$ correspond to the effective cut-offs relevant to the model. The cut-off appears with the appropriate power $n$ to ensure a dimension 4 operator for each model, while the overall coefficient $\kappa$  varies from model to model and includes numerical factors and Higgs insertions.  We see that the scalar coupling to pairs of $W$ bosons is controlled by the single parameter $\Lambda_2$, this coupling can only be set to zero by taking $\Lambda_2$ to infinity. Of the remaining three couplings to boson pairs, only one may be set to zero at a time. In our analysis we choose to focus on the symmetric benchmark point $\Lambda_1=\Lambda_2$. The values of the scalar-boson couplings for this benchmark point are shown in Table \ref{pattern-1}.  For this special point, the $\phi \gamma Z$ couplings are exactly zero and the couplings $\phi \gamma \gamma$ and $\phi ZZ$ are equal.  For this special symmetric benchmark point the branching fractions of the exotic scalar $\phi$ to vector boson pairs are the same in the  dimension 5 scalar, $SU(2)$ doublet and, dimension 7 adjoint models. These branching fractions are given in Figure \ref{bf-plots} and verified with \textsc{Madgraph 5}. While all available diboson states have branching fraction $\mathcal{O}(1)$, we see that $W$'s are dominant. Though the couplings of the scalar state to $\gamma\gamma$ and $ZZ$ pairs are equal, the phase space factor in the decay width differs between the massive and massless final states. This creates a small difference in branching fraction for lower mass $\phi$, but as $\phi$ becomes more massive, $m_{\phi} \gg m_Z$ and the $Z$'s become effectively massless compared to the heavy scalar and thus the photon and $Z$ branching fractions converge.

Models B,D and E contain scalars coupling  to $B_{\mu \nu}W^{\mu \nu}$ and  display separate coupling patterns given in Table~\ref{pattern-2}.

\begin{table}[H]  
\centering
\begin{tabular}{|c|} 
\hline 
Pattern 2 \\
\hline
\\
$V_{\phi\gamma\gamma} = -k \left[\frac{c_w s_w }{\Lambda_{\text{3}}^n} \right]$
\\
\\
$V_{\phi\gamma Z} = k \left[\frac{s_w^2 -c_w^2}{2\Lambda_{\text{3}}^n} \right]$
\\
\\
$V_{\phi ZZ} =k \left[ \frac{c_w s_w }{\Lambda_{\text{3}}^n} \right]$
\\
\\
\hline 
\end{tabular} 
\quad
\begin{tabular}{|c|}
\hline
     dimension 5 adjoint \\
     \hline
     \\
     $n=1$ $k=-1/\sqrt{2}$\\
     \\
     \hline
     weak doublet II\\
     \hline
     \\
     $n=2$ $k=v_h/2$\\
     \\
     \hline
    dimension 7 singlet\\
    \hline
    \\
     $n=3$ $k=v_h^2/4$ \\
     \\
     \hline

\end{tabular}
\caption{The effective couplings of the models B, D, and E, in terms of cutoff $\Lambda_3$ (left) and values of coefficients k and powers n for benchmark models(right)}
\label{pattern-2}
\end{table}

Here $\Lambda_3$ is the model relevant cut-off scale, the power $n$ ensures a dimension 4 term in the Lagrangian, and the overall factor k is a coefficient containing numerical factors and power of Higgs insertions for the given model. In these models, the absence of a $W_{\mu\nu}W^{\mu\nu}$ term in the Lagrangian ensures that there is no coupling between the exotic scalar $\phi$ and pairs of $W$ bosons.  The remaining three diboson couplings are controlled by a single effective cut-off parameter $\Lambda_3$, thus the relative couplings to $\gamma\gamma,\ ZZ,$ and $\gamma Z$ pairs are fixed with  $\phi \gamma \gamma$ and $\phi ZZ$ couplings being equal.  Using these couplings we can compute the  branching fractions of the scalar states with this coupling pattern, shown in Figure~\ref{bf-plots} and verified with \textsc{Madgraph 5}. We see again that with equal coupling but differing phase space factors for massless and massive particles, the di-photon branching fraction starts out the highest for low mass exotic states, but converges with the $ZZ$ branching fraction for high mass scalars. The $\gamma Z$ branching fraction is generically lower but still quite substantial.


\rev{ Explicitly, the limits on the effective mass scale are constructed as follows. First, production cross sections and scalar decay widths for each model described above are computed via \textsc{MadGraph 5}. For each computation, we use a value of 10 TeV for each finite $\Lambda_i$, and we decouple all other operators by taking their cutoffs to a value of $10^9$ GeV. Full cross sections for each considered final state are then calculated using the narrow width approximation, along with the known branching fractions of the Standard Model weak bosons \cite{Zyla:2020zbs}. For these couplings and for all masses up to 2 TeV, we find total widths less than a few percent of the mass. Therefore, we take the narrow width approximation as justified. Because the total cross section is proportional to the square of the coupling (i.e. $\Lambda^{-2n}$), cross sections for arbitrary values of $\Lambda$ can be obtained by rescaling the test result appropriately. We then mark a point as excluded if the resulting cross section exceeds the existing limit shown in Figure~\ref{fig:xs_limit_summary}. Computation of the vector boson fusion cross section using the same parameters confirms that it is orders of magnitude smaller than that of associated production.
}

We note, however, that there is a lower limit  of theoretical viability to the value of the effective operators obtained with these limits.  As a minimal bound we expect that the effective operator analysis is sensible when the effective cut-off is equal to or grater than the center of mass energy for an event.  We are considering heavy scalars, above 600 GeV in mass,  with associated gauge bosons much lighter; therefore the minimum center of mass energy in the event is approximately equal to the exotic scalar mass itself. Thus an approximate lower theoretical limit to our effective cut-offs are reached when $\Lambda \sim m_{\phi}$. 

\begin{figure}[h!]
    \centering
    \begin{subfigure}{0.45\textwidth}
        \scalebox{0.5
        }{\includegraphics{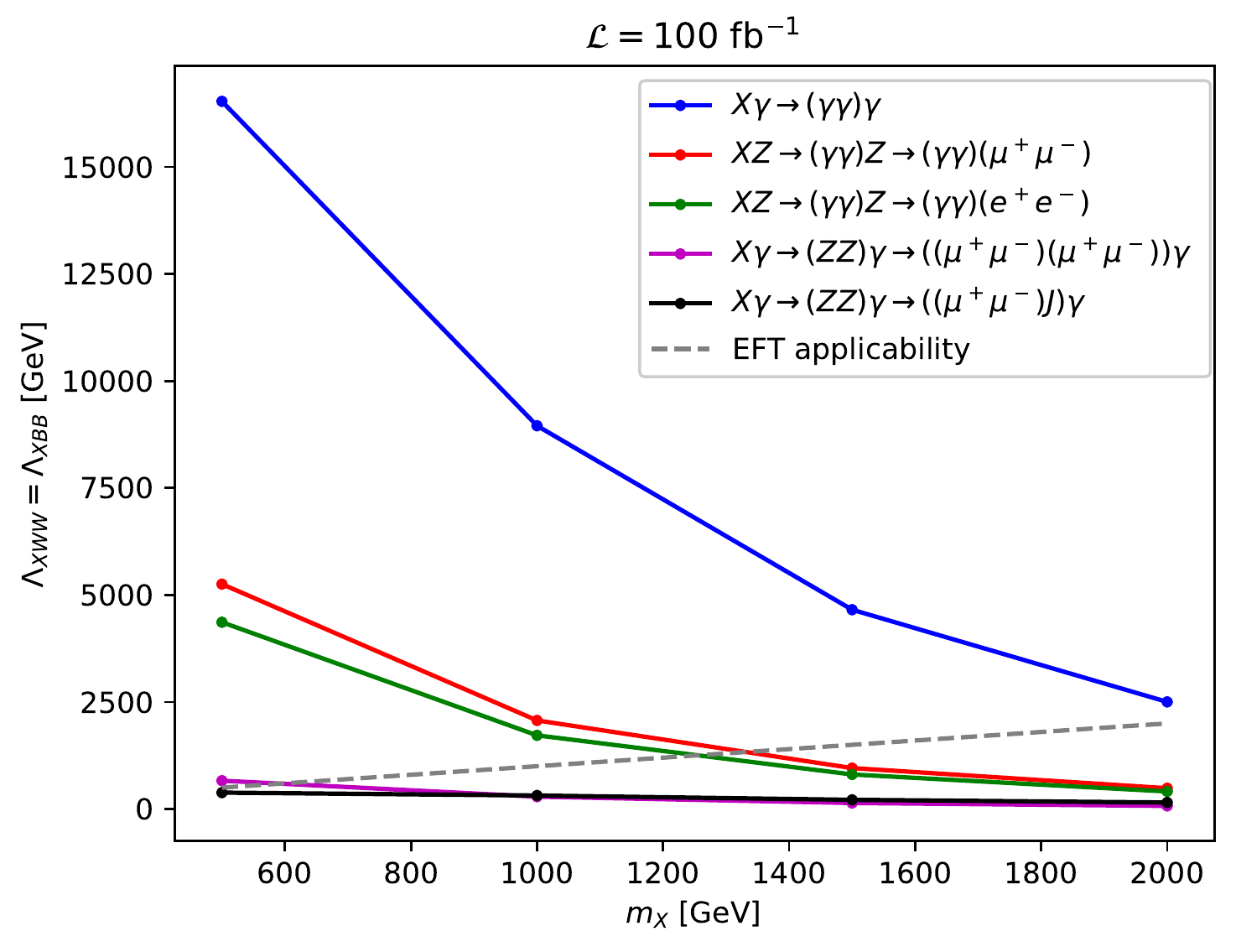}}
        \caption{dim 5 singlet 100 fb$^{-1}$}
        \label{fig:5s100}
    \end{subfigure}
    \begin{subfigure}{0.45\textwidth}
        \scalebox{0.5}{\includegraphics{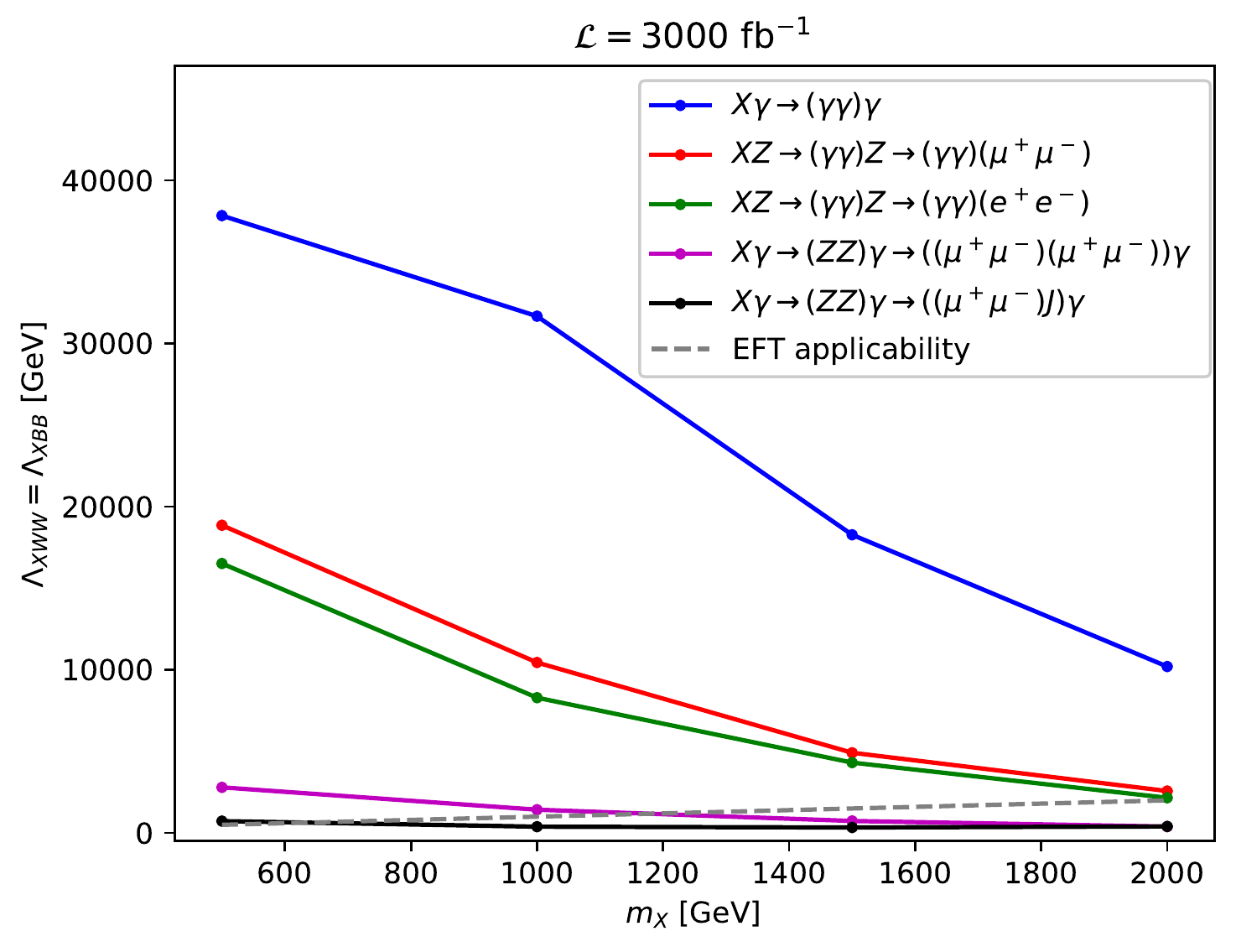}}
        \caption{dim 5 singlet 3 ab$^{-1}$}
        \label{fig:5sHL}
    \end{subfigure}
     \begin{subfigure}{0.45\textwidth}
        \scalebox{0.5}{\includegraphics{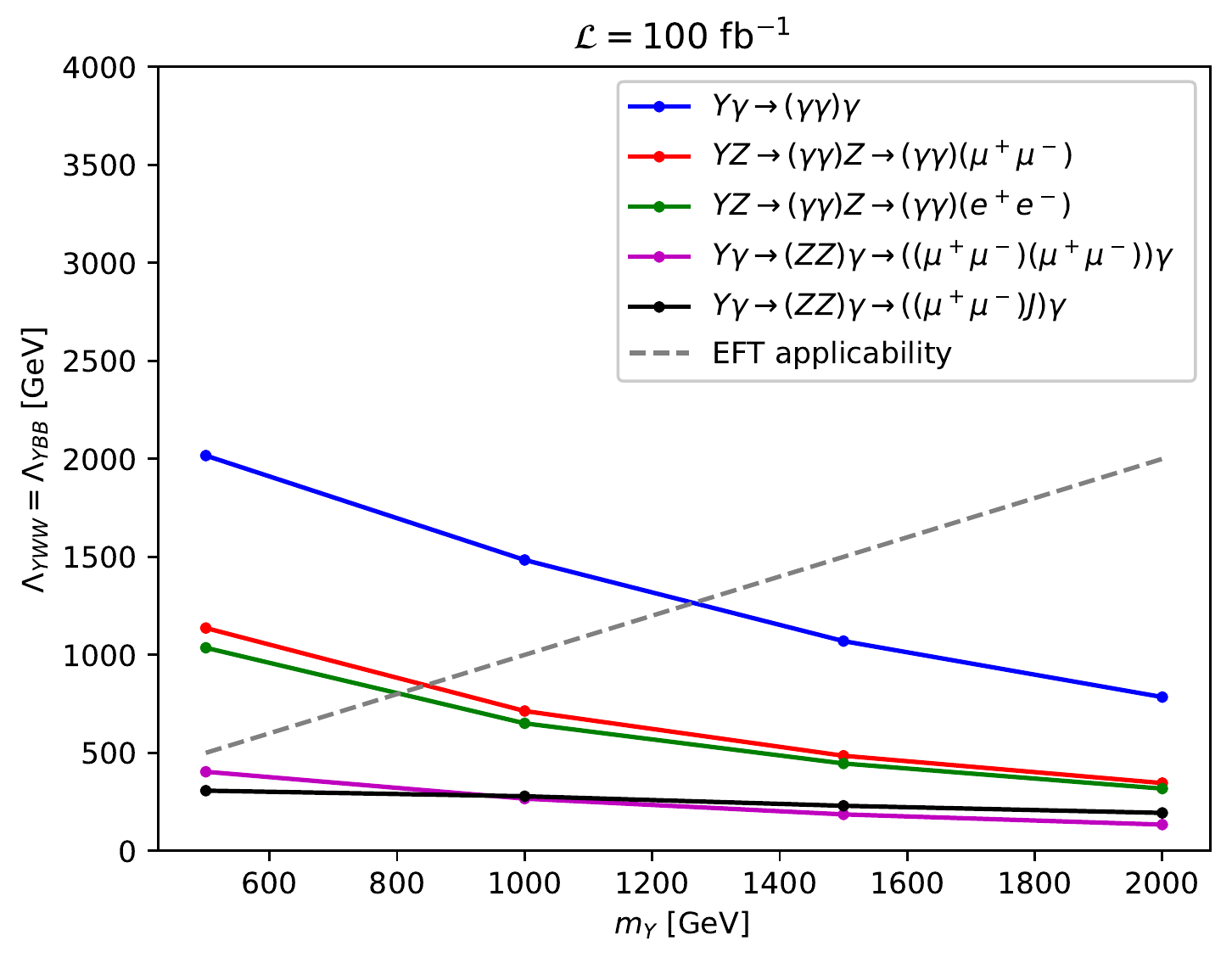}}
        \caption{dim 6 doublet model I 100 fb$^{-1}$}
        \label{fig:6I100}
    \end{subfigure}
    \begin{subfigure}{0.45\textwidth}
        \scalebox{0.5}{\includegraphics{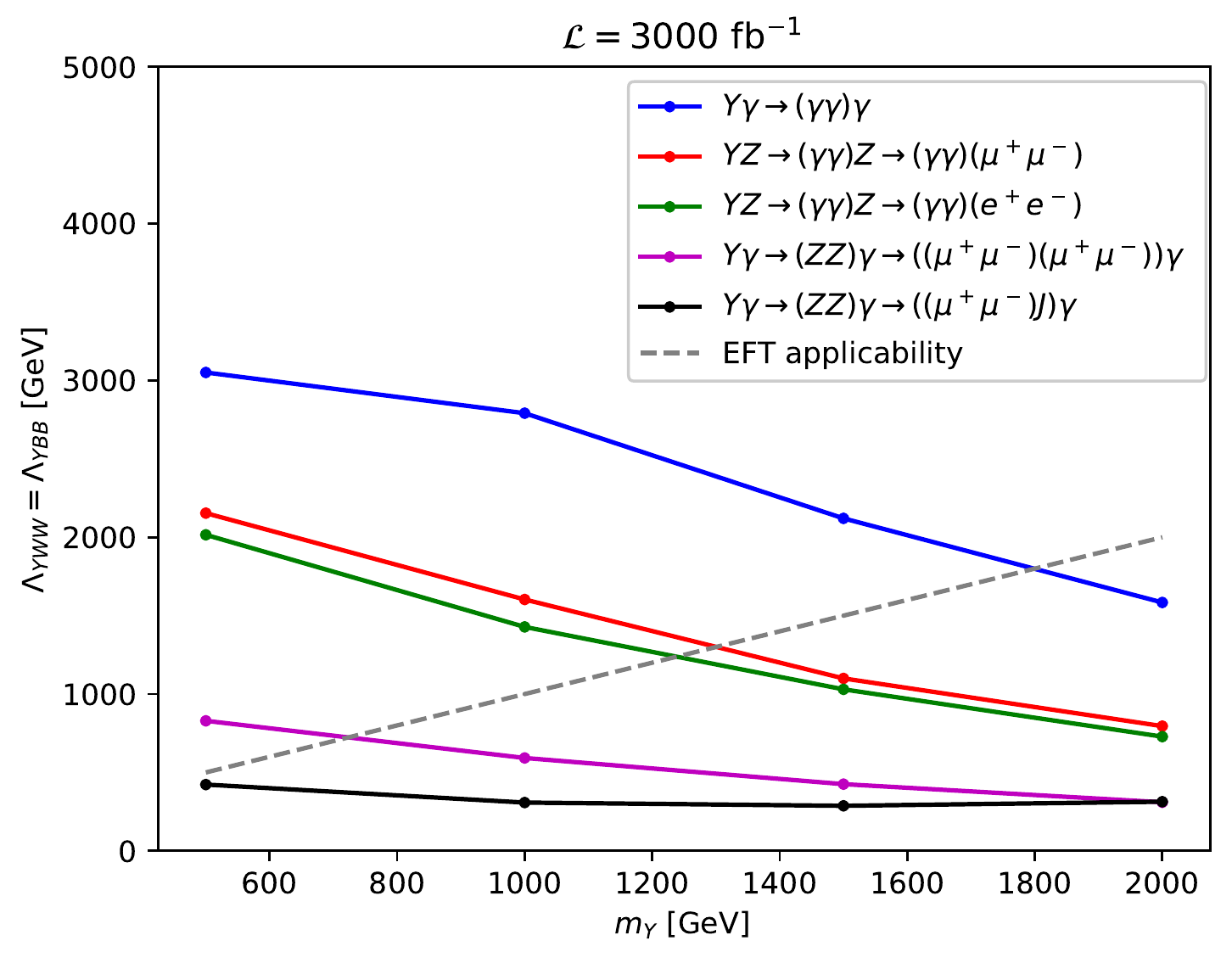}}
        \caption{dim 6 doublet model I 3 ab$^{-1}$}
        \label{fig:6IHL}
    \end{subfigure}
       \begin{subfigure}{0.45\textwidth}
        \scalebox{0.5}{\includegraphics{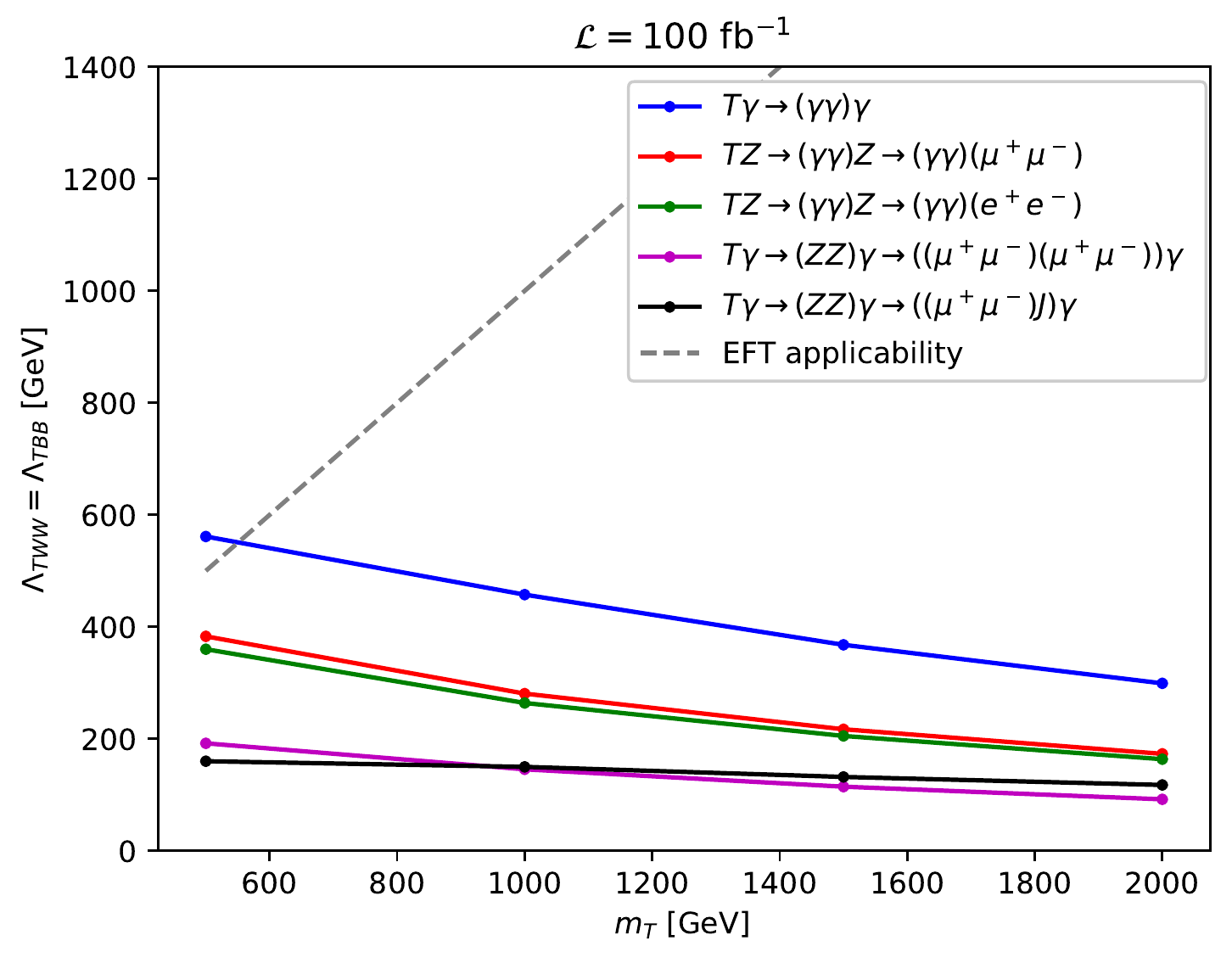}}
        \caption{dim 7 adjoint model 100 fb$^{-1}$}
        \label{fig:7a100}
    \end{subfigure}
       \begin{subfigure}{0.45\textwidth}
        \scalebox{0.5}{\includegraphics{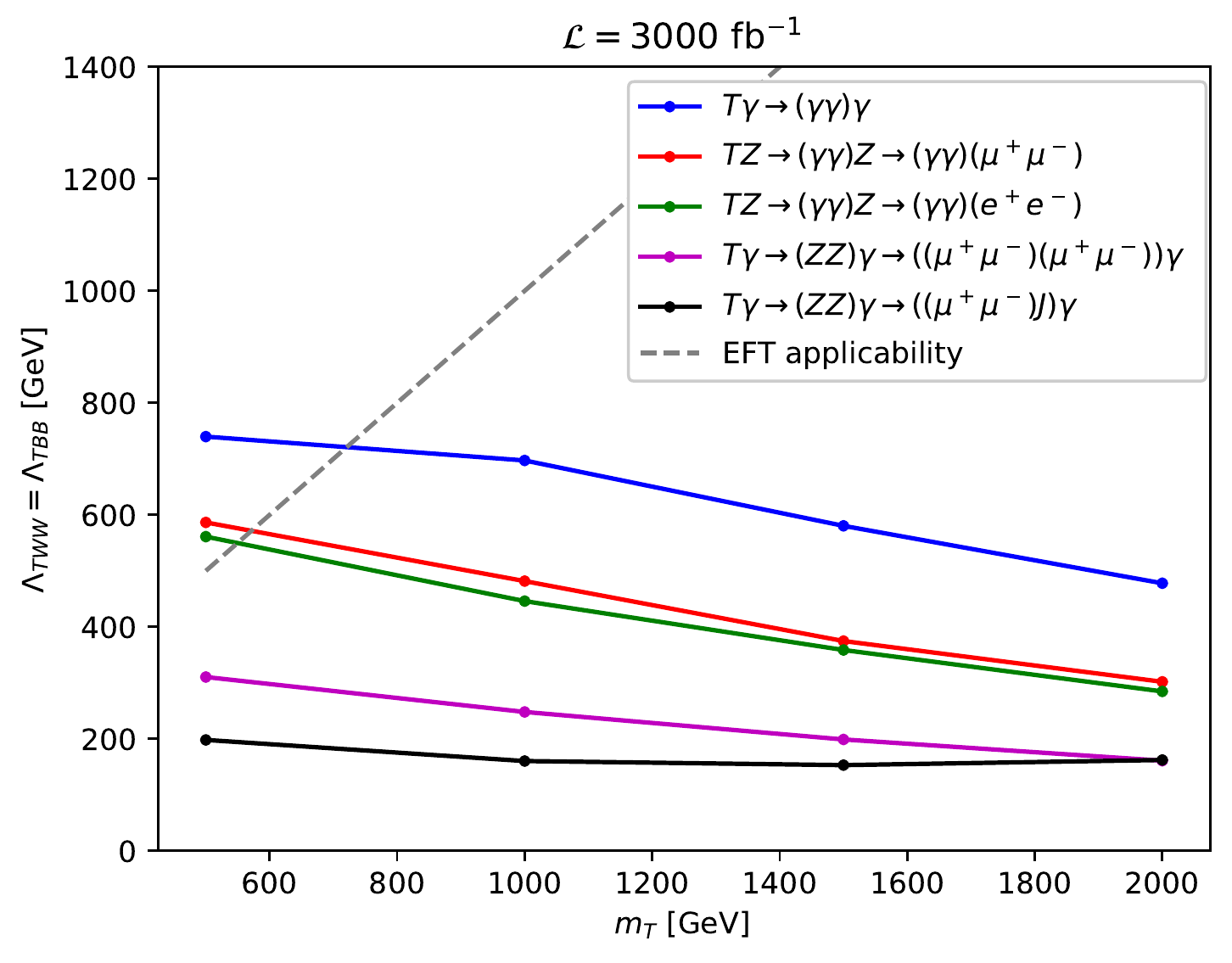}}
        \caption{dim 7 adjoint model 3 ab$^{-1}$}
        \label{fig:7aHL}
    \end{subfigure}
\caption{Exclusions of effective cut-offs vs scalar mass for 100 fb$^{-1}$ and 3 ab$^{-1}$ for dimension 5 singlet(top), dimension 6 doublet I (middle) and dimension 7 adjoint(bottom) models in various final states.}
\label{patternI}    
\end{figure}

We will now discuss expected limits in our model parameter space beginning with the Pattern 1 models; the dimension 5 singlet, doublet model I, and dimension 7 adjoint; see Figure~\ref{patternI}.   
\begin{table}
\begin{tabular}{|c|}
\hline
     $\Lambda_1=\Lambda_2=\Lambda$ \\
     \hline
     \\
     $V_{\phi \gamma\gamma}=  \frac{\kappa}{\Lambda^n}$\\
     \\
     $V_{\phi W W}= \frac{\kappa}{\Lambda^n}$\\
     \\
     $V_{\phi \gamma Z}= 0$\\
     \\
     $V_{\phi ZZ}= \frac{\kappa}{\Lambda^n}$\\
     \\
     \hline
     
\end{tabular}
\caption{The effective couplings of the models A, C, and F for the symmetric case.}
\label{pattern-1}
\end{table}

In general these models have three parameters, the mass of the singlet scalar, and the two effective cut-offs $\Lambda_{XWW}$ and $\Lambda_{XBB}$. We have analyzed results in the symmetric benchmark point  $\Lambda_{XWW} =\Lambda_{XBB}$. In this benchmark, the $\phi_{Z\gamma}$ coupling vanishes, and this fact has two phenomenological consequences. First, scalar states are produced with associated $\gamma$ or $Z$ bosons in the $s$ channel without destructive interference between photons and $Z$'s, allowing larger production cross sections. Second, the $\gamma Z$ decay channel is closed to the scalar, restricting the available final states. Of the remaining final state topologies, we have chosen to mostly reproduce here limits from channels with muons rather the electrons since the cross section limits are generally tighter, as seen in Figure~\ref{fig:xs_limit_summary}.

In Figure~\ref{patternI} we show parameter space limits for the dimension 5 singlet scalar model (top), dimension 6 weak doublet I model (middle), and dimension 7 adjoint model (bottom).  The left column shows expected limits provided no excess is seen in the current 100 fb$^{-1}$ data set, while the right column shows expected limits for the full 3 ab$^{-1}$ HL-LHC data set.  As expected, the limits on effective cut-offs weaken as we go up in effective operator dimension, as the scalar production cross-section becomes suppressed by multiple powers of the effective cut-off. The shape of the operator limit changes from low to high luminosity due to Poisson statistics. For higher mass scalars, only a small number of signal events is expected during the low luminosity run, and this allows the cross section limits to show greater improvement when the luminosity increases. This translates to a higher limit on the effective mass scale.  \rev{Also shown on each plot is a dashed line where the effective operator cutoff is equal to the scalar mass. As discussed above, this is an approximate limit of applicability for the EFT approach. Then, these plots demonstrate that the constraints on the dimension 7 operators are far less meaningful than those on the operators of lower dimension.}

\rev{For all three models, the more sensitive channels involve the reconstruction of the scalar state as a di-photon resonance. The most sensitive for each model is the tri-photon final state; here the scalar $\phi$ is produced in association with a photon, and it subsequently decays to two photons $pp \rightarrow \gamma X \rightarrow \gamma \gamma \gamma$.} The $\phi+ V$ production cross section drops quite quickly with increasing scalar mass. We might then expect the tri-photon search to become much weaker at higher scalar masses. However, the tri-photon search efficiency slightly increases with scalar mass, while the expected SM backgrounds fall substantially. We see from Figure \ref{fig:xs_limit_summary} that expected cross section limits in the tri-photon search improve as the scalar mass grows from 500 GeV to 1 TeV. The net outcome of these competing effects is that the effective cut-off exclusion falls with increasing scalar mass, but it falls by less than an order of magnitude over the 0.5-2 TeV scalar mass range.  We see that the tri-photon channel is extremely powerful in the dimension 5 singlet model, allowing exclusions of effective cut-offs between approximately 5 and 17 TeV for scalars up to 2 TeV of mass in the $100$ fb$^{-1}$ data set. Projections for 3000 fb$^{-1}$ take the limit up to 37 TeV. Limits are in the 2-3 TeV range for effective cut-offs in the dimension 6 weak doublet model.  In the case of the dimension 7 adjoint model, the production cross section is suppressed by six powers of the effective cut-off.  Therefore the tri-photon search is only marginally able to probe effective cut-offs of 500-750 GeV near the threshold of the light exotic scalar mass.

\rev{Following the tri-photon channel, the next most sensitive channels are those where the scalar is produced in association with a leptonically decaying $Z$ boson, while the scalar itself decays to a di-photon mass resonance.} The red and green lines on the plots correspond to the $\phi Z\rightarrow(\gamma\gamma)Z\rightarrow (\gamma\gamma)\mu\mu$ and $\phi Z\rightarrow(\gamma\gamma)Z\rightarrow (\gamma\gamma)ee$ channels respectively. These channels offer a complementary window into the multi-TeV scale as they have a similar order of magnitude reach as the tri-photon search. These channels also have a falling profile of cut-off sensitivity as the scalar mass increases. In the dimension 5 singlet model these channels are expected to exclude up to 5 TeV in effective cut-offs for light scalars in the 100 fb$^{-1}$ data set and up to 15-20 TeV in the 3 ab$^{-1}$ data set. In the full HL-LHC data set, the channels maintain theoretical viability in the entire range of scalar masses studied, with limits of about 5 TeV in effective cut-off for 2 TeV scalar masses.  In the dimension 6 doublet model, these channels maintain theoretical viability only for lighter scalar masses.  For example, these channel exclude 2 TeV cut-offs for light scalars in the full HL-LHC data set, but reach the $\Lambda \sim m_{\phi}$ viability threshold for scalars roughly 1.2 TeV in mass.  For dimension 7 operators,  the channel gives marginal exclusions of 600 GeV effective cut-offs for light scalars, but use of the EFT is not theoretically viable for higher scalar masses.

We consider two more eclectic channels in which the exotic scalar is produced in association with a photon and decays to two $Z$ bosons.  The most striking final states are the 4 muon and 2 muon plus large-radius jet channels, that is  $\phi\gamma\rightarrow (ZZ)\gamma \rightarrow ((\mu\mu)(\mu\mu))\gamma$ and $\phi\gamma\rightarrow (ZZ)\gamma \rightarrow ((\mu\mu) (J))\gamma$.   We can see from the exclusion plot in Figure \ref{fig:xs_limit_summary} that the four muon channel makes a slight improvement in sensitivity as the scalar mass increases.  However, the $2\mu+$ large-radius-jet channel makes a stunning increase in sensitivity, over an order of magnitude, as the scalar mass increases.  Though the overall cross section limit in the $2\mu+$ large-radius-jet channel seems worse than other channels, the large $Z$ branching fraction to jets makes it competitive with leptonic channels in the overall search.  For example, in the $4\mu$ search, each $Z$ boson must pay the price of a few percent branching fraction which cuts down on the total signal.  This makes $4\mu$ and $2\mu+$ large-radius-jet similar in reach.  

These two channels only have reach into the effective parameter space for the lightest scalars; however, the signals are so striking that they are worth mention. For example, we can see from the plots in Figure \ref{patternI} that the sensitivity of the $2 \mu+$ large-radius jet channel has the interesting feature of roughly maintaining the same exclusion power over the entire scalar mass range. This is due to the falling production cross section competing with the greatly improved efficiency and background control. The $4\mu$ channel has a hard photon and 4 leptons which reconstruct the two $Z$ bosons in the event and the heavy scalar mass resonance. We see from the figure that the HL-LHC exclusions have some sensitivity to cut-offs in the 500-700 GeV mass range in these channels while still maintaining theoretical viability of the EFT for the dimension 5 singlet and dimension 6 doublet models.

\begin{figure}[h!]
    \centering
    \begin{subfigure}{0.45\textwidth}
        \scalebox{0.5
        }{\includegraphics{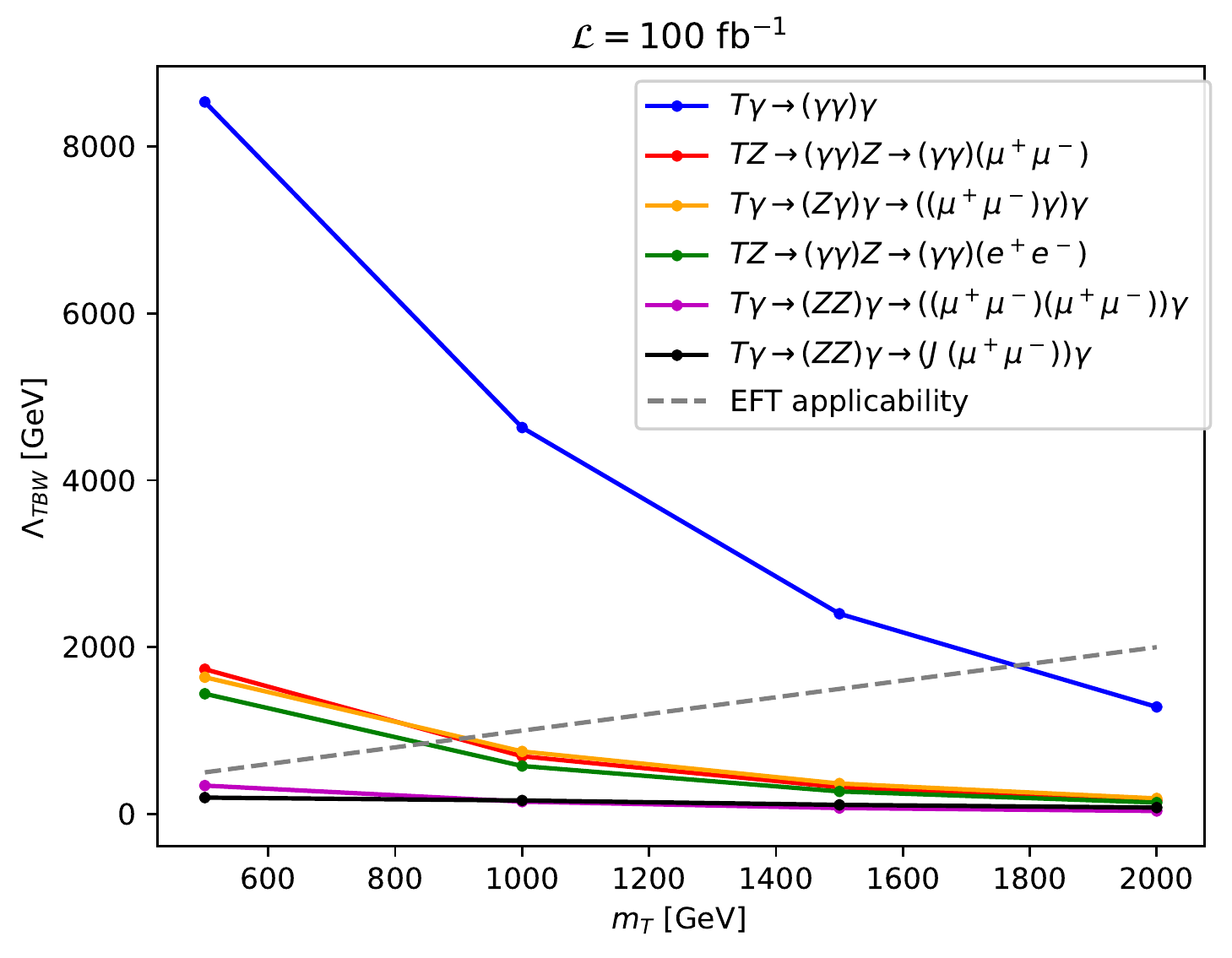}}
        \caption{dim 5 adjoint 100 fb$^{-1}$}
        \label{fig:aphi-1a4m}
    \end{subfigure}
    \begin{subfigure}{0.45\textwidth}
        \scalebox{0.5}{\includegraphics{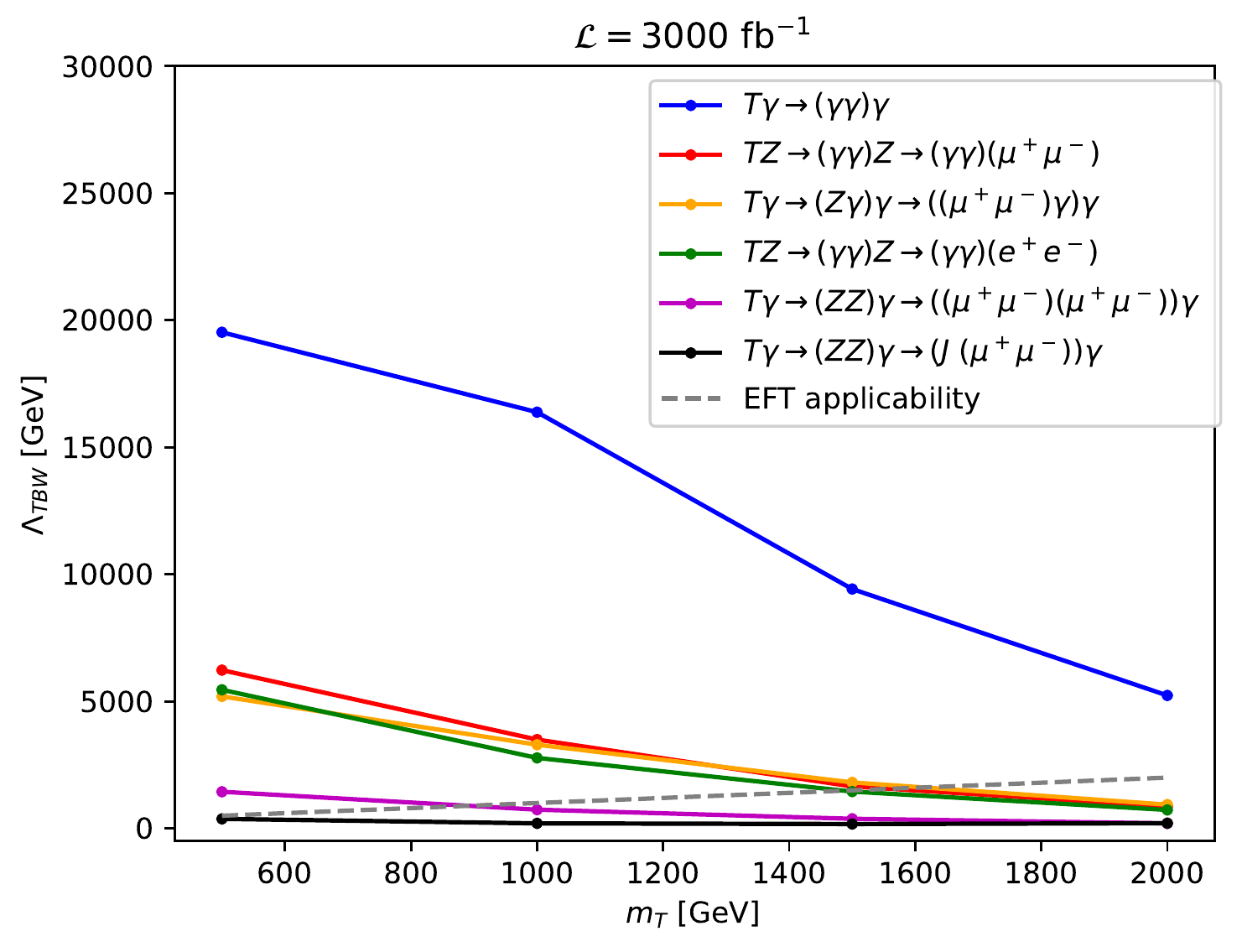}}
        \caption{dim 5 adjoint 3 ab$^{-1}$}
        \label{fig:aphi-1a4e}
    \end{subfigure}
     \begin{subfigure}{0.45\textwidth}
        \scalebox{0.5}{\includegraphics{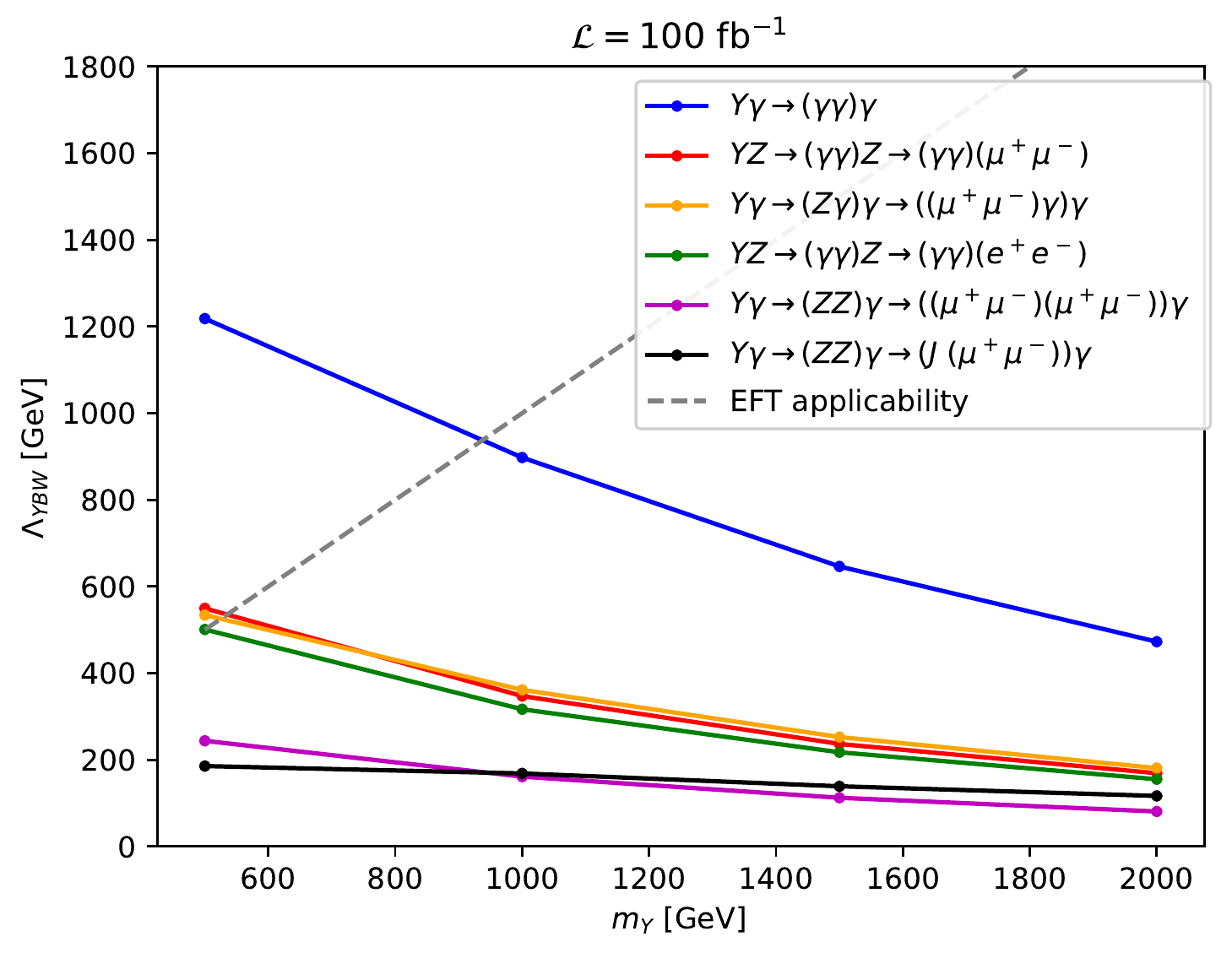}}
        \caption{dim 6 doublet model II 100 fb$^{-1}$}
        \label{fig:aphi-1a4e}
    \end{subfigure}
    \begin{subfigure}{0.45\textwidth}
        \scalebox{0.5}{\includegraphics{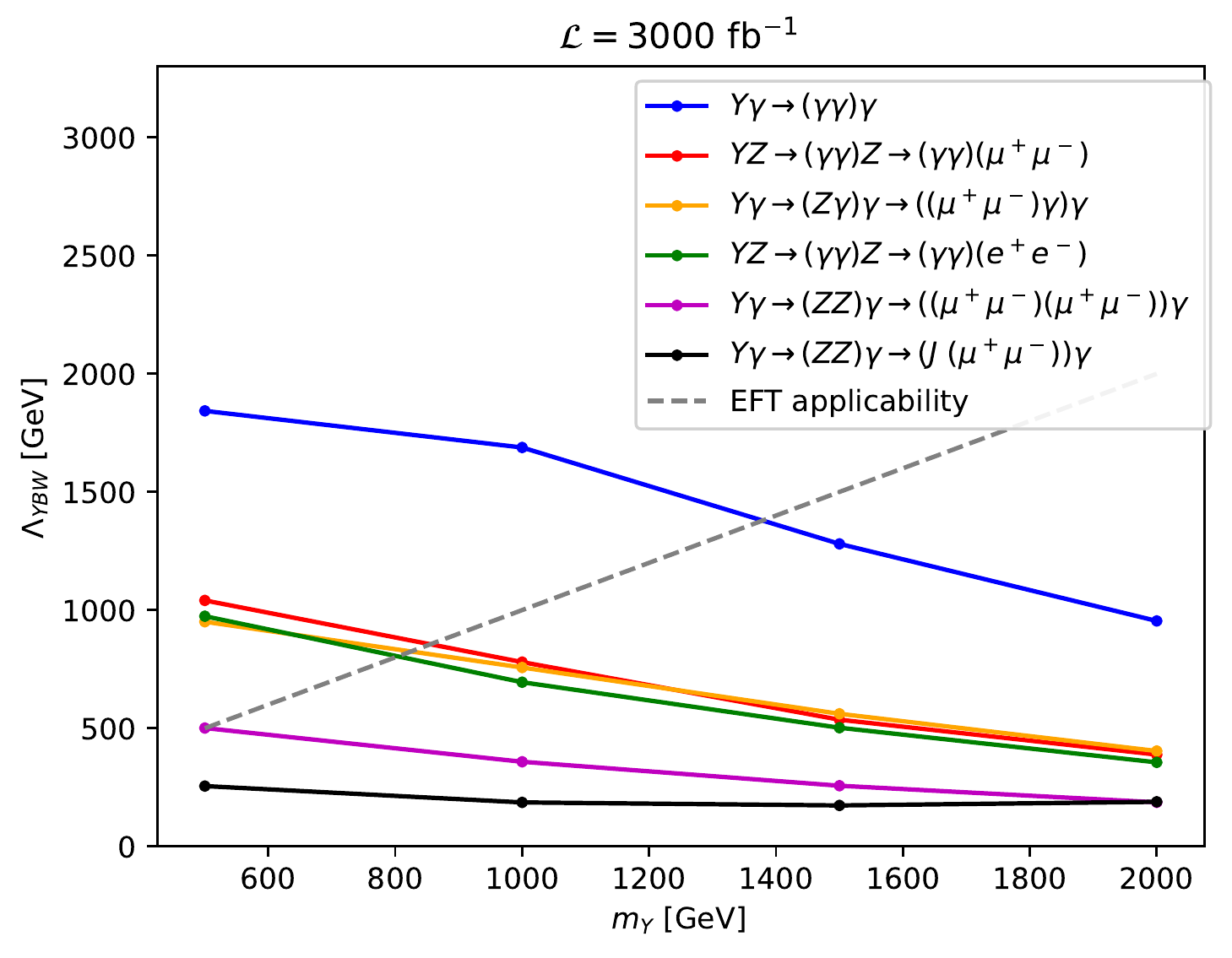}}
        \caption{dim 6 doublet model II 3 ab$^{-1}$}
        \label{fig:aphi-1a4e}
    \end{subfigure}
       \begin{subfigure}{0.45\textwidth}
        \scalebox{0.5}{\includegraphics{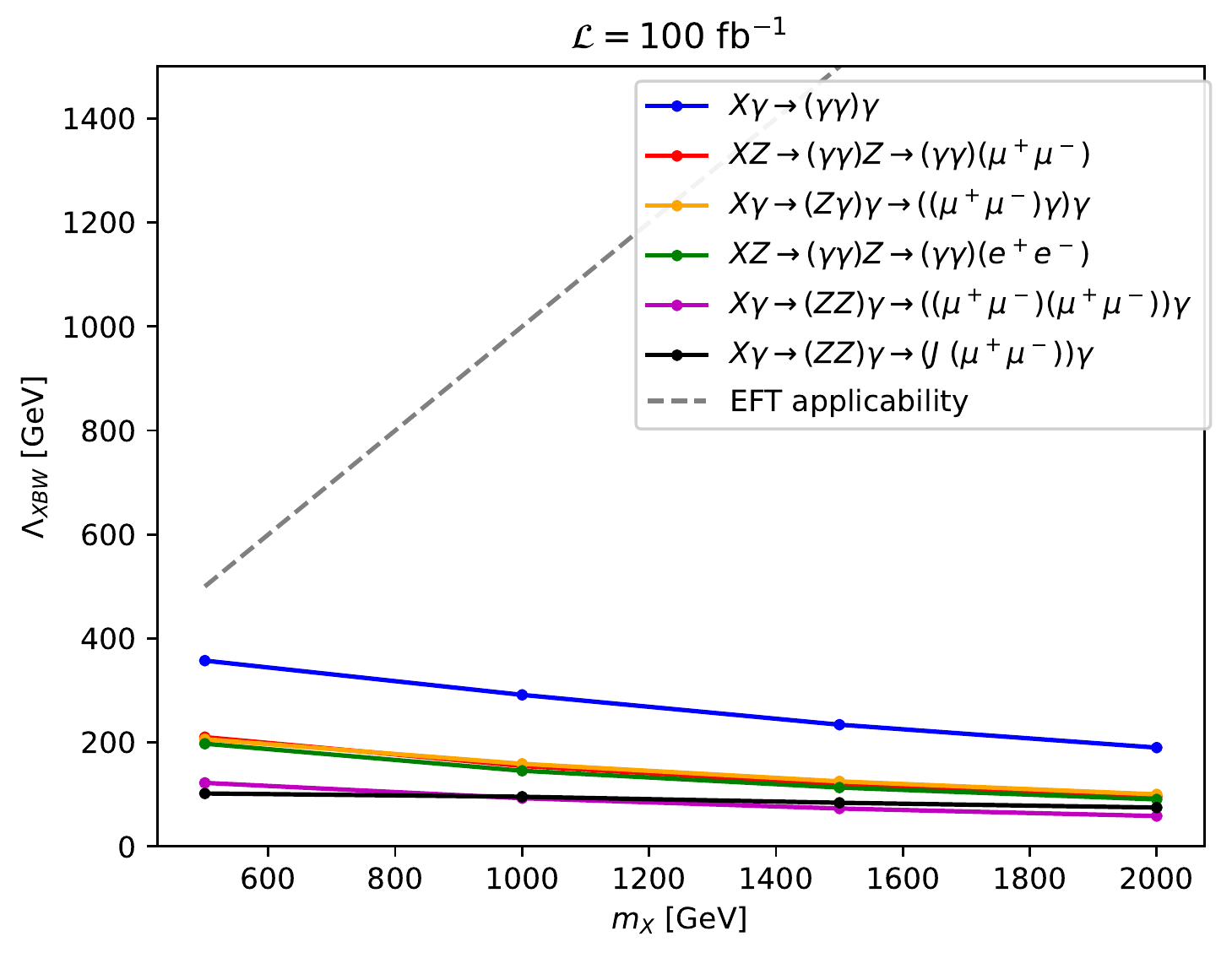}}
        \caption{dim 7 singlet model  100 fb$^{-1}$}
        \label{fig:aphi-1a4e}
    \end{subfigure}
       \begin{subfigure}{0.45\textwidth}
        \scalebox{0.5}{\includegraphics{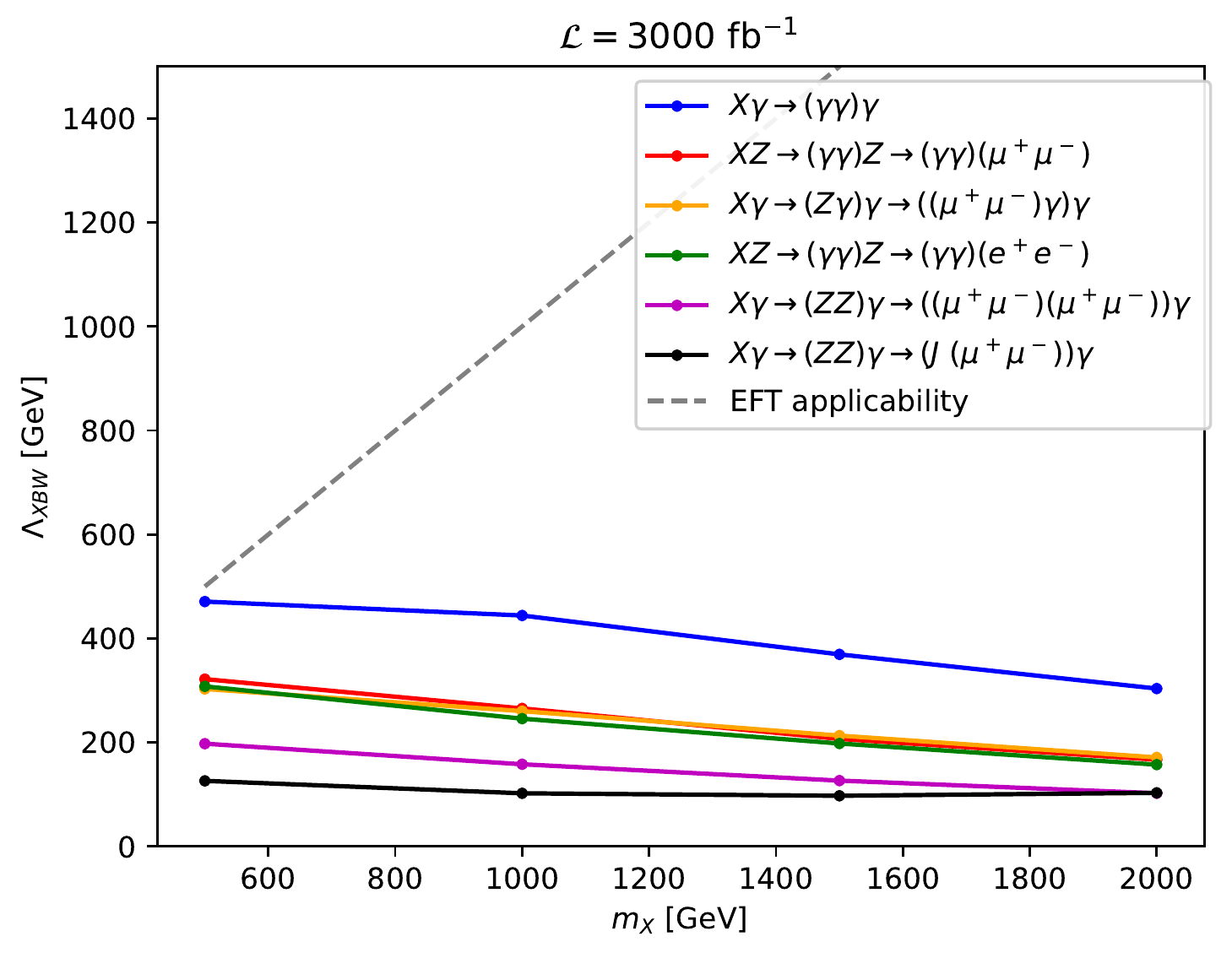}}
        \caption{dim 7 singlet model 3 ab$^{-1}$}
        \label{fig:aphi-1a4e}
    \end{subfigure}
    \caption{Exclusions of effective cut-offs vs scalar mass for 100 fb$^{-1}$ and 3 ab$^{-1}$ for dimension 7 singlet (top), dimension 6 doublet II (middle) and dimension 5 adjoint (bottom) models in various final states.}
    \label{patternII}
\end{figure}

We will now discuss the models with coupling pattern 2: the dimension 5 adjoint, weak doublet II, and dimension 7 singlet models.  We can see from Table \ref{pattern-2} that these models couple the exotic scalar $\phi$ to three sets of gauge bosons, $\gamma\gamma$, $Z\gamma$, and $ZZ$ with a single effective cut-off parameter.  The ratio of these couplings is completely fixed with respect to each other and thus no coupling may be set to zero by itself.  In these models, the $Z\gamma$ coupling is always non-zero, which has two consequences; one is that there is destructive interference between virtual gauge bosons in the $\phi+V$ production channel, and the other is that the $Z\gamma$ decay mode of the exotic scalar is always open. 

In Figure \ref{patternII} we have shown parameter space limits for the dimension 5 adjoint scalar model (top), dimension 6 weak doublet II model (middle), and dimension 7 scalar model (bottom).  Again, the left column shows expected limits provided no excess is seen in the current 100 fb$^{-1}$ data set, while the right column shows expected limits for the full 3 ab$^{-1}$ HL-LHC dta set.  

Exclusions for our final states follow much the same features as those for for the Pattern I models discussed above.  Once again the tri-photon channel -shown in blue- is the most constraining channel for all models.  We see that for the dimension 5 adjoint model effective cut-offs have an expected exclusion between 8 TeV-2 TeV over the 500 to 2 TeV scalar mass range in the 100 fb$^{-1}$ data set.  Exclusions range between 20 TeV-6 TeV over the scalar mass range for the full 3 ab$^{-1}$ HL-LHC data set. For the dimension 6 doublet model, the production cross-sections are suppressed by $v^2/\Lambda^2$ and limits are therefore weaker. Expected exclusions are in the 1-2 TeV range for light scalars under about a TeV. Moreover, the expected exclusions traverse the theoretical limit of $\Lambda \sim m_{\phi}$ for scalar masses around 900 GeV for 100 fb${}^{-1}$ of data, and for masses about 1.5 TeV in the 3 ab${}^{-1}$ data-set.  The dimension 7 singlet model has extremely suppressed production cross-sections. In this scenario only the tri-photon search can marginally probe the theoretically viable effective parameter space, excluding $\Lambda\sim 500 $GeV for scalars of the same mass scale. 

In this set of models the $Z\gamma$ decay channel of exotic scalars must be non-zero.  Therefore we have included on these plots a final state in which the scalar is produced in association with a photon and decays to $Z\gamma$; specifically the decay channel is $\phi \gamma \rightarrow  (Z\gamma)\gamma \rightarrow ((\mu\mu)\gamma) \gamma$.  This channel has similar sensitivity to the above mentioned 2 lepton- 2 photon channels and exclusions are shown in Figure \ref{patternII} in yellow.  This channel and the  $\phi Z\rightarrow(\gamma\gamma)Z\rightarrow (\gamma\gamma)\mu\mu$ (shown in red) and $\phi Z\rightarrow(\gamma\gamma)Z\rightarrow (\gamma\gamma)(ee)$ (in green) channels give a complimentary probe to the tri-photon channel into the multi-TeV effective cutoff range.  In the dimension 5 adjoint model these channels are expected to exclude cut-offs up to 2 TeV in 100 fb${}^{-1}$ of data and up to 7 TeV in 3 ab${}^{-1}$ of data for light scalar masses.  The limit of theoretical viability is traversed in the 100 fb${}^{-1}$ data set for scalar masses above about 1 TeV, however the search remains viable over almost the entire range of adjoints masses in the full HL-LHC data set. The dimension 6 doublet model fairs a bit less well, though these channels are expected to exclude up to 1 TeV of effective cutoff for light scalars in the full 3 ab${}^{-1}$ data set-the limit of theoretical viability is traversed even in the full data set for scalar masses of about 800 GeV.  This channel can marginally exclude cut-offs of 500-600 GeV for 500 GeV scalars with 100 fb${}^{-1}$ of data. These channels do not have reach into the theoretically viable region of parameter space for the dimension 7 singlet model.

The $\phi\gamma\rightarrow (ZZ)\gamma \rightarrow ((\mu\mu)(\mu\mu))\gamma$ (shown in purple) is expected to have some reach for light scalars in the 3 ab${}^{-1}$ data set.  In the dimension 5 adjoint model it can probe about 1 TeV in effective cut-off for 500 GeV masses, and is theoretically viable up to about 800 GeV in scalar mass. It can also marginally probe 500 GeV cut-offs for scalar of similar mass in the weak doublet model.  In these scenarios the $\phi\gamma\rightarrow (ZZ)\gamma \rightarrow ((\mu\mu) (J))\gamma$ channels unfortunately fail to reach the threshold of theoretical viability.

\section{Conclusions}

We have introduced the LHC search topology of triple electroweak gauge bosons to study the  production of exotic states that couple to pairs of electroweak gauge bosons. The exotic particles are produced in association with a photon or $Z$ boson and decay to electroweak gauge bosons pairs.  We have studied 14 possible final states with this topology. We have shown how the presence of multi-photons, multiple leptons, and mass reconstruction of both $Z$ bosons and heavy exotic scalars in the event can give extremely tight constraints on new models, possible placing cross section limits in the sub-fb region. 

We have proposed a set of new effective field theory models where exotic scalar states  in various representations of SM gauge groups couple to pairs of SM gauge bosons.  We have analyzed our results in the parameter space of the effective cut-off scales of the EFTs.  We find several channels capable of probing the multi-TeV effective cut-off regime, with the powerful tri-photon searches reaching as far as 35 TeV in effective cut-off for the full 3  ab$^{-1}$ run of HL-LHC.  

There are several  directions for future work. First, we could imagine that more sensitivity may be gained by combining channels studied in this work. One simple one might include the analysis of the triple electroweak boson channels with associated $W$ bosons. This would extend the study to the production of exotic charged scalar states produced in association with a $W$.  Another direction may be to extend these results to more models, perhaps by extending the EFT catalogue or mapping the results to fleshed out models, for example the sgaugino sector of R symmetric supersymmetry.  Yet another direction may be to explore unusual final state topologies resultant from the diboson couplings of the EFT models catalogued in this work.

\section{Acknowledgments}
LMC and MJS were funded in part by the United States Department of Energy under grant DE-SC0011726. DW and JC were funded by the DOE Office of Science.

\section{Appendix A: Catalogue of scalar couplings}
For the sake of completeness we explicitly write the exotic scalar couplings to sets of electroweak gauge boson in our 6 benchmark models.

\begin{equation}
\begin{split}
&\mathrm{Model\ A: dimension\ 5\ singlet\ } X 
\\[2ex]
&\ \ \ \ V_{X\gamma \gamma} =  \frac{ c_w^2} {\Lambda_{\text{XBB}}}+\frac{ s_w^2 }{\Lambda_{\text{XWW}}}
\\[1.5ex]
&\ \ \ \ V_{XWW} =  \frac{1}{\Lambda_{\text{XWW}}}   
\\[1.5ex]
&\ \ \ \ V_{X\gamma Z} = -\frac{c_w s_w}{\Lambda _{\text{XBB}}}+\frac{ c_w s_w}{\Lambda _{\text{XWW}}}  
\\[1.5ex]
&\ \ \ \ V_{XZZ} = \frac{ c_w^2}{\Lambda _{\text{XWW}}} +\frac{ s_w^2}{\Lambda _{\text{XBB}}}
\end{split}
\end{equation}

\begin{equation}
\begin{split}
&\mathrm{Model\ B: dimension\ 7\ singlet\ } X
\\[2ex]
&\ \ \ \ V_{X\gamma\gamma} = -\frac{c_w s_w v^2}{4\Lambda _{\text{XBW}}^3} 
\\[1.5ex]
&\ \ \ \ V_{X\gamma Z} = \frac{s_w^2 v^2-c_w^2v^2}{8\Lambda _{\text{XBW}}^3} 
\\[1.5ex]
&\ \ \ \ V_{XZZ} = \frac{c_w s_w v^2}{4\Lambda _{\text{XBW}}^3} 
\end{split}
\end{equation}

\begin{equation}
\begin{split}
&\mathrm{Model\ C: dimension\ 6\ doublet\ (I)\ } Y 
\\[2ex]
&\ \ \ \ V_{Y\gamma \gamma} =  \frac{ c_w^2 v} {\Lambda_{\text{YBB}}^2}+\frac{ s_w^2 v}{\Lambda_{\text{YWW}}^2}
\\[1.5ex]
&\ \ \ \ V_{YWW} =  \frac{v}{\Lambda_{\text{YWW}}^2}   
\\[1.5ex]
&\ \ \ \ V_{Y\gamma Z} = -\frac{c_w s_w v}{\Lambda _{\text{YBB}}^2}+\frac{ c_w s_w v}{\Lambda _{\text{YWW}}^2}  
\\[1.5ex]
&\ \ \ \ V_{YZZ} = \frac{ c_w^2 v}{\Lambda _{\text{YWW}}^2} +\frac{ s_w^2 v}{\Lambda_{\text{YBB}}^2}
\end{split}
\end{equation}

\begin{equation}
\begin{split}
&\mathrm{Model\ D: dimension\ 6\ doublet\ (II)\ } Y
\\[2ex]
&\ \ \ \ V_{Y\gamma\gamma} = -\frac{c_w s_w v^2}{2\Lambda_{\text{YBW}}^2}
\\[1.5ex]
&\ \ \ \ V_{Y\gamma Z} = \frac{s_w^2 v^2-c_w^2 v}{4\Lambda_{\text{YBW}}^2}
\\[1.5ex]
&\ \ \ \ V_{YZZ} = \frac{c_w s_w v}{2\Lambda_{\text{YBW}}^2} 
\end{split}
\end{equation}

\begin{equation}
\begin{split}
&\mathrm{Model\ E: dimension\ 5\ triplet\ } T 
\\[2ex]
&\ \ \ \ V_{T\gamma \gamma} = \frac{c_w s_w }{\sqrt{2}\Lambda_{\text{TBW}}} 
\\[1.5ex]
&\ \ \ \ V_{T\gamma Z} = \frac{c_w^2 -s_w^2 }{2\sqrt{2} \Lambda_{\text{TBW}}}  
\\[1.5ex]
&\ \ \ \ V_{TZZ} = -\frac{c_w s_w}{\sqrt{2} \Lambda_{\text{TBW}}}
\end{split}
\end{equation}

\begin{equation}
\begin{split}
&\mathrm{Model\ F: dimension\ 7\ triplet\ } T
\\[2ex]
&\ \ \ \ V_{T\gamma\gamma} = -\frac{ c_w^2 v^2} {4\sqrt{2}\Lambda_{\text{TBB}}^3} -\frac{ s_w^2 v^2 }{4 \sqrt{2} \Lambda_{\text{TWW}}^3}
\\[1.5ex]
&\ \ \ \ V_{TWW} = -\frac{v^2}{4 \sqrt{2} \Lambda_{TWW}^3}
\\[1.5ex]
&\ \ \ \ V_{T\gamma Z} = \frac{c_w s_w v^2}{ 4 \sqrt{2} \Lambda _{\text{TBB}}^3}-\frac{ c_w s_w v^2}{4 \sqrt{2} \Lambda _{\text{TWW}}^3}
\\[1.5ex]
&\ \ \ \ V_{TZZ} = -\frac{ c_w^2 v^2}{ 4 \sqrt{2} \Lambda _{\text{TWW}}^3} -\frac{ s_w^2 v^2}{4 \sqrt{2} \Lambda _{\text{TBB}}^3}
\end{split}
\end{equation}

\clearpage

\bibliography{tribosons}

\end{document}